\documentclass[10pt,a4paper]{article}
\usepackage{amsmath,fourier,amssymb,amsthm,graphicx,epstopdf,color}
\usepackage{chngcntr}
\counterwithout{figure}{section}
\usepackage[T1]{fontenc}
 \theoremstyle{plain}
 \newtheorem {hypo}{\bf\hspace{-\parindent}Hypothesis}

 \newtheorem {lemma}[hypo]{Lemma}

 \theoremstyle{remark}
 \newtheorem {rmk}[hypo]{Remark}
  
 \newcommand{\pf}{\begin{bpf}}

 \newcommand{\pfms}{\begin{bpfms}}
 \newcommand{\epf}{\end{bpf}\hfill$\square$\vspace{0.1cm}}
 \newcommand{\epfms}{\end{bpfms}\hfill$\square$\\ }
 \newcommand\ben{\begin{equation*}}
 \newcommand\ebn{\end{equation*}}
 \newcommand\beq{\begin{equation}}
 \newcommand\eeq{\end{equation}}
 
  \newcommand\lb{\left(}
  \newcommand\rb{\right)}

\begin{document}
	
 \LARGE
\noindent
\textbf{On some Hamiltonian properties of the isomonodromic \vspace{0.1cm}\\ 
tau functions}
\normalsize
 \vspace{1cm}\\
 \textit{
 A. R. Its \footnote{aits@iupui.edu} , 
A. Prokhorov\footnote{aprokhor@iupui.edu}} 
 \vspace{0.2cm}\\
 Department of Mathematical Sciences,
 Indiana University-Purdue University,
 402 N. Bla\-ckford St.,\\
 Indianapolis, IN 46202-3267,
 USA.
\vspace{0.2cm}\\
St. Petersburg State University, Universitetskaya emb. 7/9, 199034, St. Petersburg,
Russia.

\begin{abstract}
We discuss some new aspects of the theory of the Jimbo-Miwa-Ueno tau function which  have come to light within the recent
developments in the global asymptotic analysis of the  tau functions related to the Painlev\'e equations. Specifically, we show that
up to the total differentials the logarithmic derivatives of the  Painlev\'e tau functions coincide with the  corresponding classical action differential.
This fact simplifies considerably the evaluation of the constant factors in the  asymptotics of tau-functions, which has been
a long-standing problem of the asymptotic theory of Painlev\'e equations. Furthermore, we believe that this observation is yet another manifestation
of L. D. Faddeev's emphasis of the key role which the Hamiltonian aspects play in the theory of integrable system. 
\end{abstract}
\vspace{0.1cm}
\begin{center}
	{\it This article will appear in the WSPC memorial volume dedicated to Ludvig 
		Faddeev.}
\end{center}

\section{Introduction}
Consider a system of linear ordinary differential equations with rational coefficients,
  \beq\label{gensys0} 
 \frac{d\Phi}{dz}= A\lb z\rb\Phi,
 \eeq
  where $ A\lb z\rb$ is an $N\times N$, $N>1$ matrix-valued rational function,
  The object of our study is the   Jimbo-Miwa-Ueno   tau function associated with the isomonodromic deformation of system (\ref{gensys0}).
  Let us remind, following \cite{JMU}, the general set-up associated with this notion.

 Denote the poles of the matrix valued rational  function $ A\lb z\rb$ on ${\mathbb C}P^1$ by $a_1,\ldots ,a_n,\infty$ and by $r_1, r_2, ..., r_n, r_{\infty}$ the corresponding
 Poincar\'e  ranks. The matrix function $A(z)$ can be then written as,
 \beq\label{A}
 A\lb z\rb =\sum_{\nu=1}^{n}\sum_{k=1}^{r_\nu+1}
 \frac{A_{\nu,-k+1}}{\lb z-a_\nu\rb^{k}}+\sum_{k=0}^{r_\infty-1}z^kA_{\infty,-k-1},\quad A_{\nu,-k+1},\,\,A_{\infty,-k-1}\in \mathfrak{sl}_N\left(\mathbb C\right), \qquad k=1,\ldots,r_\nu+1, \quad \nu=1,\ldots,n. 
\nonumber \eeq
  We are going to make the standard assumption that all highest order matrix coefficients $A_{\nu}\equiv A_{\nu,-r_\nu}$ are diagonalizable
 \ben
 A_{\nu,-r_{\nu}}=G_{\nu}\Theta_{\nu,-r_\nu}G_{\nu}^{-1};\quad \Theta_{\nu,-r_\nu}=\operatorname{diag}\left\{\theta_{\nu,1},\ldots \theta_{\nu,N}\right\},
 \ebn
  and that their eigenvalues are distinct and non-resonant: 
 \ben
 \begin{cases}
 	\theta_{\nu,\alpha}\neq \theta_{\nu,\beta} \quad &\mbox{if}\quad r_\nu \geq 1,\quad \alpha\neq \beta,\\
 	\theta_{\nu,\alpha}\neq \theta_{\nu,\beta} \mod \mathbb{Z}\quad &\mbox{if}  \quad r_{\nu}=0,\quad \alpha\neq \beta.
 \end{cases}
 \ebn
At each singular point, the system (\ref{gensys0}) admits a unique {\it formal} solution,
 \beq\label{formalsol}
 \Phi_{\mathrm{form}}^{(\nu)}\left(z\right) =  G^{\lb\nu\rb}\left(z\right)e^{\Theta_{\nu}\lb
 z\rb}, \qquad \nu = 1, \ldots , n,\,\infty,
\eeq
where $G^{\lb\nu\rb}\left(z\right)$ are formal series,
$$
G^{\lb\nu\rb}\left(z\right) = G_{\nu}\left[ I+\sum_{k=1}^{\infty}g_{\nu,k}\left(z-a_{\nu}\right)^k\right],
\quad
G^{\lb\infty\rb}\left(z\right) = G_{\infty}\left[  I+\sum_{k=1}^{\infty}g_{\infty,k}z^k\right],
$$
and  $\Theta_{\nu}(z)$ are diagonal matrix-valued functions,
$$
\Theta_{\nu}(z)=\sum_{k=-r_\nu}^{-1}
\frac{\Theta_{\nu,k}}{k}\lb z-a_\nu\rb^{k}+\Theta_{\nu,0}\ln \lb z-a_\nu\rb,\qquad
\Theta_{\infty}\lb z\rb =-\sum_{k=1}^{r_{\infty}}\frac{\Theta_{\infty,-k}}{k}z^{k}-\Theta_{\infty,0}\ln z.
$$
For every $\nu\in\{1, \ldots, n,\infty\}$,  the matrix coefficients $g_{\nu,k}$ and $\Theta_{\nu,k}$ can be explicitly computed  in terms 
of the coefficients of the matrix-valued rational function $G^{-1}_{\nu} A\lb z \rb G_{\nu}$, see \cite{JMU}.

 The non-formal global properties of solutions of the equation (\ref{gensys0}) are described by its
 {\it monodromy data} $M$ which include: i) formal monodromy exponents $\Theta_{\nu,0}$, ii) appropriate connection
 matrices between canonical solutions at different singular points, and iii) relevant Stokes matrices at irregular ($r_{\nu} \geq 1$) singular
 points. Let us denote the space of monodromy data of the system (\ref{gensys0}) by $ \mathcal{M} $. It can be described  in more
details as follows.
 
 
 Let $a_{\nu}$ be an irregular singular point of index $r_{\nu}$. For $j=1,\ldots, 2r_{\nu}+1$, let also
 \begin{equation}\label{stokssec1}\nonumber
 \Omega_{j,\nu} = \left\{z :  0<|z-a_{\nu}|<\epsilon,\quad \theta^{(1)}_j< \arg \lb z-a_{\nu}\rb < \theta^{(2)}_{j},
 \quad \theta^{(2)}_j - \theta^{(1)}_j = \frac{\pi}{r_{\nu}} + \delta \right\}
 \end{equation}
 be the {\it Stokes sectors} around $a_{\nu}$ (see, e.g., \cite[Chapter 1]{FIKN} or \cite{Was} for more details). According to the general 
 theory of linear systems, in each sector $ \Omega_{j,\nu}$ there exists a unique {\it canonical solution}  $\Phi_j^{(\nu)}\left(z\right)$ of (\ref{gensys0}) which satisfies the asymptotic condition 
 \begin{equation}\label{can1}\nonumber
  \Phi_j^{(\nu)}\left(z\right)\simeq \Phi^{(\nu)}_{\mathrm{form}}\lb z\rb \qquad\text{as } z\to a_{\nu},\quad  z \in \Omega_{j,\nu},\quad j=1,\ldots, 2r_\nu +1.
  \end{equation}
 Different canonical solutions are related by {\it Stokes matrices}, $S_j^{(\nu)}$,  and {\it connection matrices}, $C_{\nu}$:
 \ben
 \Phi_{j+1}^{(\nu)}=\Phi_{j}^{(\nu)}S_j^{(\nu)},\quad j = 1,\ldots, 2r_{\nu}, \qquad  \Phi_{1}^{(\nu)}=\Phi_{1}^{(\infty)}C_{\nu},\quad
 \nu = 1,\ldots, n.
 \ebn  
 Let us assume  that the irregular singular points are $\infty$ and the first $n_0\leq n$ points among the singular points $a_1, \ldots, a_n$. Denote by ${\mathcal S}_{\nu}$ the collection of Stokes matrices at an irregular point $a_{\nu}$, i.e. 
  \begin{equation}\label{sphen}\nonumber
  {\mathcal S}_{\nu} =\left\{S_1^{(\nu)}, \ldots, S_{2r_{\nu}}^{(\nu)}\right\}.
  \end{equation}
 The space ${\mathcal M}$ of monodromy data of the system (\ref{gensys0}) consists of formal monodromy exponents $\Theta_{\nu,0}$,  connection matrices $C_{\nu}$ and Stokes matrices $S_j^{\lb \nu\rb}$, i.e.,
 \begin{equation}\label{mset}\nonumber
 {\mathcal M} = \left\{ M \equiv \Bigl(\Theta_{\nu, 0},\,\, \nu = 1, \ldots, n, \infty; \quad  C_{\nu},\,\,  \nu = 1,\ldots, n;\quad  
 {\mathcal S}_{\nu},\,\,  \nu = 1,\ldots, n_0,  \infty\Bigr) \right\}.
 \end{equation}
 We shall use the notation,
  $$
    \vec{m} = \lb m_1, \ldots, m_d\rb, \quad d = N(n+1) + nN^2+ \left(\dfrac{N(N-1)}{2}\right)\lb\sum_{\nu=1}^{n_0}2r_{\nu} + 2r_{\infty}\rb,
$$
for the points $ \vec{m} \in{\mathcal M}$.
In addition, we    denote  by $\mathcal{T}$ the set of {\it times}, 
 \begin{equation}\label{istime1}
 a_{1}, \ldots, a_{n}, \quad  (\Theta_{\nu,k})_{ll}, \quad k= -r_\nu,\ldots,-1,\quad \nu = 1, \ldots, n_0,\, \infty,
 \quad l = 1, \ldots, N.\nonumber
 \end{equation}
We shall use the notation,
 $$
 \vec{t} = \lb t_1, \ldots, t_L\rb, \quad L = n + N\lb\sum_{\nu=1}^{n_0}r_{\nu} + r_{\infty}\rb,
 $$
 for the points $\vec t\in\mathcal T$.  Let us also denote by $\mathcal{A}$ the variety of all rational matrix-valued 
functions $ A \lb z\rb$ with a fixed number of poles  of fixed orders.
The so-called Riemann-Hilbert correspondence states that, up to  submanifolds where 
the inverse monodromy problem for (\ref{gensys0}) is not solvable, the space $ \mathcal{A}$ can be identified with
the product $\widetilde{\mathcal T}\times \mathcal{M}$, where $\widetilde{\mathcal T}$ denotes the universal covering of $\mathcal{T}$. We shall loosely write,
$$
\mathcal{A} \simeq \widetilde{\mathcal T}\times \mathcal{M}.
$$
It should be mentioned that in each concrete case one has to specify the gauge normalization of the matrix $A(z)$ as well as
the choice of the gauge matrices $G_{\nu}$ in order to make this identification  well defined. In Section \ref{3} we will demonstrate how
these specifications can be done in the case of  Painlev\'e equations.

The {\it Jimbo-Miwa-Ueno 1-form}  is defined as the following differential form on $\mathcal{A}$:
\begin{equation}\label{jmu}
\omega_{\mathrm{JMU}}=-\sum_{\nu=1,\ldots, n, \infty} \mathop{\mathrm{res}}_{z=a_\nu} \mbox{Tr}\left(\left(G^{(\nu)}
\lb z\rb\right)^{-1}\frac{dG^{(\nu)}}{dz}\lb z\rb\, d_{\mathcal T}\Theta_{\nu}\lb z\rb\right),
\end{equation}
where $G^{(\nu)}(z)$ are the series from \eqref{formalsol} and we put $a_{\infty} \equiv \infty$. The notation $d_{\mathcal T}\Theta_{\nu}\lb z\rb$ stands for
$$
d_{\mathcal T}\Theta_{\nu}\lb z\rb = \sum_{k=1}^{L}\frac{\partial\Theta_{\nu}\lb z\rb}{\partial t_{k}}\,dt_k,\qquad 
L = n + N\left(\sum_{\nu=1}^{n_0}r_{\nu} + r_{\infty}\right).
$$
The significance of this
form is that, being restricted to any {\it isomonodromic family} in the space $\mathcal{A}$,
$$
A\lb z\rb \equiv  A\lb z; \vec{t},  M\rb,\qquad \vec{t} = \lb t_1,\ldots, t_L\rb,\quad M \equiv \mbox{const}
$$
it  becomes closed with respect to times $\mathcal T$, i.e.
\begin{equation}\label{jmu2}\nonumber
d_{\mathcal T}\left(\omega_{\mathrm{JMU}}\bigl|_{ A\lb z; \vec{t}, M\equiv \mathrm{const}\rb}\right) = 0.
\end{equation}
The closedness of the 1-form $\omega_{\mathrm{JMU}}$ with respect to $\mathcal T$ in turn implies that locally there is a function $\tau \equiv \tau\lb\vec{t};M\rb$ on ${\mathcal{T}}
\times \mathcal M$ such that
\begin{equation}\label{jmu3}
d_{\mathcal T}\ln \tau = \omega_{\mathrm{JMU}}\bigl|_{ A\lb z; \vec{t}, M\rb}.
\end{equation}
A remarkable property of the {\it tau function}  $\tau\lb\vec{t}, M\rb$, which was established in   \cite{Malgrange} and \cite{Miwa}, is that it 
admits analytic continuation  as an entire function to the whole universal covering $\widetilde{\mathcal T}$ of the parameter space ${\mathcal T}$.
Furthermore, zeros of $\tau\lb\vec{t}, M\rb$ correspond to the points in ${\mathcal T}$ where the inverse monodromy problem
for (\ref{gensys0}) is not solvable for a given set $M$ of monodromy data  (or, equivalently, where the holomorphic vector
bundle over ${\mathbb C}P^1$ determined by $ M$ becomes nontrivial). Hence a central role of the concept of tau function in
the monodromy theory of systems of linear differential equations.

The isomonodromicity of the family $ A\lb z; \vec{t}, M\rb$ means
that all equations from it have the same set $ M\in \mathcal{M}$ of monodromy data. This  implies that the corresponding solution   
$\Phi\lb z\rb \equiv \Phi\lb z,\vec{t}\rb$  
 satisfies an overdetermined system 
 \beq
 \begin{cases} \label{isosys}
 \,\,\partial_z\Phi\,\, =  A\lb z,\vec t\rb \Phi\lb z,\vec{t}\rb,\\
 d_{\mathcal{T}}\Phi =B\lb z,\vec t\rb\Phi\lb z,\vec{t}\rb.
 \end{cases}
 \eeq
 The coefficients of the matrix-valued differential form $B \equiv \sum_{k=1}^{L}B_k\lb z,\vec t\rb dt_k$ are rational in $z$. Their explicit form may be algorithmically deduced from the expression for $A\lb z\rb$ (see again \cite{JMU}). The compatibility of the system (\ref{isosys}) yields the {\it  monodromy preserving deformation equation}:
 \begin{equation}\label{isomeg0}
 {d_{\mathcal{T}}A}=\partial_z B+[B,A].
 \end{equation}
 Isomonodromy equation (\ref{isomeg0}) is of great interest on  its own. Indeed, it includes as special cases practically
all known integrable differential equations. The first nontrivial cases of (\ref{isomeg0}), where the set of isomonodromic times 
effectively reduces to a single variable $t$,  cover all six classical Painlev\'e equations. Solutions of the latter are dubbed as {\it nonlinear special functions},
and they indeed play this role in many areas  of modern nonlinear science (see \cite{FIKN}, \cite{BK},\cite{DS},\cite{GM},
\cite{TW1},\cite{TW2}).

The principal  analytic issue concerning the tau function, in particular from the point of view of applications, is its  behavior near the critical hyperplanes, where either $a_{\mu} = a_{\nu}$ for some $\mu \neq \nu$, or $\theta_{\nu,\alpha} = \theta_{\nu,\beta}$ for some~$\nu$ and some $\alpha \neq \beta$. In the case of Painlev\'e equations this is the behavior of 
respective tau functions near the $t =\infty$ (PI, II, IV),  $t =\infty, 0$ (PIII, V), and $t =\infty, 0, 1$ (PVI). A special challenge in the asymptotic analysis of the 
tau functions  is the evaluation  of the constant pre-factors in their asymptotics.  In fact, it is these pre-factors which usually contain the most important information
about the physical properties of the model under investigation. At the same time, they can not be obtained directly via the Riemann-Hilbert approach. The latter method
is one of the principal modern tools of the asymptotic analysis of Painlev\'e transcendents, and it is based on the asymptotic evaluation of the above mentioned Riemann-Hilbert correspondence. In other words, the Riemann-Hilbert technique allows to evaluate the asymptotics of the matrix 
$A\lb z,\vec t\rb $
and hence the asymptotics of the differential form $\omega_{\mathrm{JMU}}$. In view of (\ref{jmu3}), this gives the asymptotics of the logarithmic derivatives of the
tau function. In order to obtain the complete asymptotic description of the tau function itself, which would include the above mentioned pre-factors, one has 
to solve the ``constant problem'':  to find the constant of integration arising from the formal integration of (\ref{jmu3}). More precisely, since the tau function is itself defined up to a multiplicative constant, we are actually  talking about the evaluation, in terms of monodromy data,  of the ratios of constant factors corresponding to different critical points (Painlev\'e~ III, V, VI) or to different critical directions (Painlev\'e I, II, IV). 

The first rigorous solution of a constant  problem for Painlev\'e equations (a special Painlev\'e III transcendent appearing in the Ising model)
has been obtained in the work of C. Tracy \cite{Tracy}. After that, several other important special cases have been also solved.  
We refer the reader to \cite{ILP} for a detailed history of the question. It is important to emphasize that all these works were concerned
with the very special families of the Painlev\'e functions, and they used the techniques which could not be extended to the generic tau functions.

The  means to solve the ``constant problem'' for   tau functions  corresponding to the generic solutions of Painlev\'e equations
started to develop  since the 2013-2014 works \cite{ILT13}, \cite{ILST} of Iorgov, Lisovyy, Shchechkin, and Tykhyy where a very important discovery of the conformal block interpretation
of tau functions was made. For the history of the question, we refer the reader to the paper \cite{ILP} where 
the heuristic  though truly pioneering results of \cite{ILT13}, \cite{ILST}  have been rigorously proven. Another conjectural 
pre-factor formula, this time concerning the third Painlev\'e equation (work \cite{ILT14}), was proven in \cite{IP}. Later on,
the method of \cite{ILP} and \cite{IP} was successfully applied to the first Painlev\'e equation in \cite{LR}.

The method of \cite{ILP} and \cite{IP} is   inspired by the earlier works  of B.  Malgrange \cite{Malgrange} and Bertola \cite{Bertola} and it is based on an extension 
of the Jimbo-Miwa-Ueno   differential form $\omega_{\mathrm{JMU}}$
 to a 1-form, $\omega$ on the whole space ${\mathcal A}\simeq \widetilde{\mathcal T}\times  \mathcal{M}$,
 $$
 \omega = \sum_{k=1}^{L}P_k(\vec{t}, M)dt_k +  \sum_{j=1}^{d}Q_j(\vec{t}, M)dm_j,
 $$
 such that 
 $$
 \omega\lb\partial_{t_k}\rb = \omega_{\mathrm{JMU}}\lb \partial_{t_k}\rb,
 $$
  and the exterior differential of $\omega$, i.e., the form,
 \beq\label{Omega0def}
 \Omega_0:= d\omega  
 \eeq
  is a 2-form on $\mathcal M$ only.  Furthermore, it is independent of isomonodromic times ${\mathcal T}$. The construction of the form  $\omega$
 will be described in detail in the next section.
\footnote{ The exact relation of the form $\omega$ to the original Malgrange-Bertola 1-form is explained in detail in \cite{ILP} -see Remark 4.4 there.} 

 The time-independence of the 2-form $\Omega_0$ in conjunction with the Riemann-Hilbert computability of the  asymptotics of $\Phi\lb z\rb$  determines what should be added to the form $\omega$ to make it closed, i.e. to transform it into the form $\hat{\omega}$ which satisfies the two crucial properties:
 $$
d \hat{\omega} \equiv  d_{\mathcal T}\hat{\omega} + d_{\mathcal M}\hat{\omega} = 0, \quad\mbox{and}\quad 
\hat{\omega}\lb\partial_{t_k}\rb = \omega_{\mathrm{JMU}}\lb \partial_{t_k}\rb.
$$
Having the form $\hat{\omega}$, the tau function can be represented as 
\begin{equation}\label{exttau0}
\ln\tau = \int \hat{\omega}.
 \end{equation}
 Equation (\ref{exttau0}) allows one to use the asymptotic behavior  of $\Phi\lb z\rb$ to evaluate the 
 asymptotics of the associated tau function  up to a numerical (i.e. independent of monodromy data) constant. The latter can be calculated  by applying the final formulae to trivial solutions of deformation equations. 
This program  has been first realized in \cite{IP}   for the sine-gordon reduction of the Painlev\'e III equation and later on  in \cite{ILP} for   Painlev\'e VI and II equations and in \cite{LR} for Painlev\'e I equation. 

We want also mention the most recent
work \cite{GL} where a general Fredholm determinant formula was found for the Painlev\'e VI tau function which allows 
to produce  rigorously both the evaluation of the relevant asymptotic constants and the combinatorial series expansions of the tau function
at the critical points. 
 
In the course of the asymptotic analysis performed in \cite{IP} and \cite{ILP}, an interesting observation has been made  with respect  to the form $\omega$ in the
Painlev\'e III and II cases.
This observation is concerned with the Hamiltonian aspect of the theory of isomonodromic deformations   which we have not yet discussed. As a matter of fact,  
the space  $\mathcal{A}$ can be equipped with a  symplectic structure - see  \cite{H2}, \cite{Boalch}, \cite{Kri}, \cite{B},  so that the isomonodromic equation (\ref{isomeg0})   induces $L$ commuting Hamiltonian flows on $\mathcal{A}$.  A striking property of the tau function is that in many
(though not all{\footnote{ The statement actually  depends on the specific choice of the symplectic structure.}})  known special cases  its  logarithm  serves as the {\it generating function} of the Hamiltonians  $H_k$ of these flows:
\begin{equation}\label{ham0}
\frac{\partial \ln\tau\lb\vec{t}, M\rb}{\partial t_k} = \gamma{H_k}\bigl|_{ A\lb z; \vec{t}, M\rb}.
\end{equation}
Here $\gamma$ is numerical constant (in many cases, $\gamma$ =1).  This fact for the fourth, fifth and sixth Painlev\'e equations
as well as for many higher rank isomonodromic systems was established in  \cite{Boalch2} where also a generalization of the
JMU form allowing repeated eigenvalues was worked out. The Hamiltonian formalism for all six Painlev\'e equations
was first suggested by  K. Okamoto \cite{O}.

The above mentioned observation of \cite{IP} and \cite{ILP}  is that in the Painlev\'e II and III cases  the 2 -form $\Omega_0 = d\omega$ is nothing else but, up to a numerical factor, the corresponding symplectic form. Hence, in these examples, the 1-form $\omega$, up to a numerical factor and  the addition of an explicit total differential, is an extension to the space 
$\widetilde{\mathcal T}\times \mathcal{M}$ of the differential of  classical
action; moreover, in \cite{IP} and \cite{ILP}  these total differentials have been explicitly found. Similar relation to the classical action in the case
of the Painlev\'e I tau-function has been obtained in \cite{LR}, and in the case of the system associated to Schlesinger equations -- the pure Fuchsian 
setup (\ref{gensys0}), in \cite{Malg2}{\footnote{ The authors are grateful to  Marta Mazzocco for pointing out this result of B. Malgrange to us.}}.

The goal of this paper is to show that the relation between the tau function and the classical action established  
in \cite{IP}, \cite{ILP}, and \cite{LR} for the special cases of Painlev\'e III, II, and I  is true for all Painlev\'e equations. 
We shall also present some arguments allowing one to expect that this relation is, most likely,  a general fact of the monodromy theory of linear systems. 

The detailed construction of the form $\omega$ is given, following \cite{ILP}, in the next section. In this section we also provide the arguments 
in favor of the  connection between the form $\omega$ and the classical action in the general  case of linear system (\ref{gensys0}) and formulate
two conjectures concerning with this connection. In Section \ref{3}, these conjectures are justified for all six  Painlev\'e equations and
for system associated to Schlesinger equations. In the cases of Painlev\'e I and Schlesinger equations we just reproduce the results of \cite{LR} and \cite{Malg2},
respectively.

\section{The extended Jimbo-Miwa-Ueno differential and the  classical action functional.}\label{2}

As in introduction, we shall,  unless the otherwise is explicitly indicated, be treating  all the objects which  are defined 
on $\mathcal{A} \simeq\widetilde{\mathcal T}\times  \mathcal{M}$  as functions of $(\vec{t},  M) \equiv (\vec{t}, \vec{m})$ .
In particular, 
$$
\omega_{\mathrm{JMU}} \equiv\omega_{\mathrm{JMU}}( \vec{t}, M) = \omega_{\mathrm{JMU}}\bigl|_{ A\lb z; \vec{t}, M\rb},
$$
and for any function $F$ on $\mathcal A$, the partial derivatives with respect to $t_k$ will mean the partial derivatives 
of $F$ as a function of $(\vec{t},  M)$, i.e.,
$$
\frac{\partial F}{\partial t_k}  \equiv \frac{\partial }{\partial t_k}F(\vec{t}, M) = \frac{\partial }{\partial t_k}\left(F\bigl|_{ A\lb z; \vec{t}, M\rb}\right).
$$
We will also use  the notations
\beq\label{not}\nonumber
dF \equiv dF(\vec{t}, M) = \sum_{k=1}^{L}\frac{\partial F }{\partial t_k}dt_k + \sum_{k=1}^{d}\frac{\partial F }{\partial m_k}dm_k
\equiv d_{\mathcal T}F + d_{\mathcal M}F.
\eeq
Our starting point is the following Lemma.
\begin{lemma}([JMU])\label{lemjmu} The $1$-form (\ref{jmu}) (considered
as a 1-form on $\widetilde{\mathcal T}\times \mathcal{M}$) can be alternatively written as
	\beq\label{ojmu2}
	\omega_{\mathrm{JMU}}=
	\sum_{k=1}^L\sum_{\nu=1,\ldots,n,\infty}\operatorname{res}_{z=a_{\nu}}
	\operatorname{Tr}\lb 
	A \,\dfrac{\partial G^{(\nu)}}{\partial t_k} \,
	{\left(G^{(\nu)}\right)}^{-1}\rb dt_k \equiv \sum_{\nu=1,\ldots,n,\infty}\operatorname{res}_{z=a_{\nu}}
	\operatorname{Tr}\lb 
	A \,d_{\mathcal{T}}G^{(\nu)} \,
	{\left(G^{(\nu)}\right)}^{-1}\rb.
	\eeq
\end{lemma}
\noindent
We give the version \cite{ILP} of the   proof of this Lemma  in Section \ref{appendix1} of the Appendix. 

A direct corollary of Lemma \ref{lemjmu} is the following integral formula for the tau function,
\beq\label{tau}
\ln \tau\equiv \ln \tau(\vec{t_1}, \vec{t_2}, M) = \intop_{\vec{t_1}}^{\vec{t_2}} \sum_{\nu,k} \mathrm{res}_{z=a_\nu} \mathrm{Tr}\left(A\dfrac{\partial G^{(\nu)}}{\partial t_k} \left(G^{(\nu)}\right)^{-1}\right) dt_k.
\eeq
Another consequence of the Lemma is the idea to take formula \eqref{ojmu2} as the motivation  to introduce  the following 1-form (cf. \cite{ILP}, \cite{IP})
$$
	\omega=
	\sum_{k=1}^L\sum_{\nu=1,\ldots,n,\infty}\operatorname{res}_{z=a_{\nu}}
	\operatorname{Tr}\lb 
	A \,\dfrac{\partial G^{(\nu)}}{\partial t_k} \,
\left(G^{(\nu)}\right)^{-1}\rb dt_k+\sum_{k=1}^d\sum_{\nu=1,\ldots,n,\infty}\operatorname{res}_{z=a_{\nu}}
	\operatorname{Tr}\lb 
	A \,\dfrac{\partial G^{(\nu)}}{\partial m_k} \,
\left(G^{(\nu)}\right)^{-1}\rb dm_k
$$
$$
\equiv\sum_{\nu=1,\ldots,n,\infty}\operatorname{res}_{z=a_{\nu}}
	\operatorname{Tr}\lb 
	A \,d_{\mathcal{T}}G^{(\nu)}\,
\left(G^{(\nu)}\right)^{-1}\rb +\sum_{\nu=1,\ldots,n,\infty}\operatorname{res}_{z=a_{\nu}}
	\operatorname{Tr}\lb 
	A \,d_{\mathcal{M}}G^{(\nu)} \,
\left(G^{(\nu)}\right)^{-1}\rb
$$ 
\begin{equation} \label{mb}
\equiv \sum_{\nu=1,\ldots,n,\infty}\operatorname{res}_{z=a_{\nu}}
	\operatorname{Tr}\lb 
	A \,dG^{(\nu)}\,
\left(G^{(\nu)}\right)^{-1}\rb.
\end{equation}
Now, the key observation.
\begin{lemma}[ILP]\label{lemma2} The form $d\omega$ has no cross terms of the kind $dt_k\wedge dm_j,\quad k=1,\ldots,L,\quad j=1,\ldots,d.$
\end{lemma}
\noindent
We present, following \cite{ILP}, the proof of this Lemma in section \ref{appendix2} of the Appendix.

Lemma \ref{lemma2} plays a crucial role in the mentioned in the Introduction rigorous approach to the ``constant problem''. Indeed, a key issue in the determining of the monodromy dependence of the  tau function is the possibility of the effective evaluation of the derivative of the integral (\ref{tau}) 
with respect to the monodromy parameters $m_j$. Lemma \ref{lemma2} implies that 
$$
\frac{\partial }{\partial m_{j}} 
 \sum_{\nu} \mathrm{res}_{z=a_\nu} \mathrm{Tr}\left(A\dfrac{\partial G^{(\nu)}}{\partial t_k}\left(G^{(\nu)}\right)^{-1}\right)
 = \frac{\partial}{\partial t_k } \sum_{\nu} \mathrm{res}_{z=a_\nu} \mathrm{Tr}\left(A{ \dfrac{\partial G^{(\nu)}}{\partial m_j}  }\left(G^{(\nu)}\right)^{-1}\right).
$$
Therefore,

$$
\frac{ \partial \ln \tau}{\partial m_j} =\intop_{\vec{t_1}}^{\vec{t_2}}  \sum_{k=1}^L\dfrac
{\partial }{\partial m_j}\sum_{\nu} \mathrm{res}_{z=a_\nu} \mathrm{Tr}\left(A\dfrac{\partial G^{(\nu)}}{\partial t_k}\left(G^{(\nu)}\right)^{-1}\right) dt_k
$$
\begin{equation}\label{action0}
=\sum_{k=1}^L\intop_{\vec{t_1}}^{\vec{t_2}}\dfrac
{\partial }{\partial t_k} \sum_{\nu} \mathrm{res}_{z=a_\nu} \mathrm{Tr}\left(A{ \dfrac{\partial G^{(\nu)}}{\partial m_j}  }\left(G^{(\nu)}\right)^{-1}\right) dt_k=\left.\sum_{\nu} \mathrm{res}_{a_\nu} \mathrm{Tr}\left(A{\dfrac{\partial G^{(\nu)}}{\partial m_j}}\left(G^{(\nu)}\right)^{-1}\right)\right|_{\vec{t_1}}^{\vec{t_2}}.
\end{equation}
In other words, we conclude that in addition to the differential relation (\ref{jmu3}), i.e.,
\begin{equation}\label{jmu33}\nonumber
d_{\mathcal T}\ln \tau = \sum_{\nu=1,\ldots,n,\infty}\mbox{res}_{z=a_{\nu}}
 \mbox{Tr}\lb 
 {G^{\lb \nu\rb}\lb z\rb}^{-1}
 A\lb z\rb \,d_{\mathcal T}G^{\lb \nu\rb}\lb z\rb \rb, 
 \end{equation}
the tau function satisfies the differential relation,
\begin{equation}\label{jmu333}\nonumber
d_{\mathcal M}\ln \tau = \sum_{\nu=1,\ldots,n,\infty}\mbox{res}_{z=a_{\nu}}
 \mbox{Tr}\lb 
 {G^{\lb \nu\rb}\lb z\rb}^{-1}
 A\lb z\rb \,d_{\mathcal M}G^{\lb \nu\rb}\lb z\rb \rb. 
 \end{equation}
These two differential identities allow to evaluate the asymptotic connection formulae up to the numerical constants and this is what
is effectively done in \cite{IP}, \cite{ILP}.

The arguments which led to the representation  (\ref{action0}) for the logarithmic derivative of the tau function with respect to $m_j$ are reminiscent 
to the variational equations for the classical action. Let us assume that we can identify the classical Darboux coordinates
{\footnote{ The Darboux coordinates on the phase spaces $\mathcal A$ corresponding  to  Painlev\'e  equations are
introduced in \cite{Harnad}, \cite{B}; the  Darboux coordinates for more general cases of the isomonodromic deformation equations are considered in \cite{H2}.}}, $p_j$, $q_j$
on the space $\mathcal A$ so that the isomonodromic deformation equations (\ref{isomeg0}) can be written as the commuting system of
Hamiltonian dynamical equations,{\footnote{ In the special case of Painlev\'e  and system associated to Schlesinger equations, which are our principal concern,  their Hamiltonian representations 
(\ref{action1}) is described  in all details in the main body of the paper. 
 The  interested reader can be referred to section 5 of \cite{Boalch2} for the general definition of a time-dependent Hamiltonian system in the
 context of isomonodromy setting. We notice that  this general definition is a delicate issue since  the original parametrization of the extended phase  
 space $\mathcal A$ mixes the time and dynamical parameters. }}
\beq\label{action1}
\frac{\partial q_j}{\partial t_k} = \frac{\partial H_k}{\partial p_j}, \quad \frac{\partial p_j}{\partial t_k} = -\frac{\partial H_k}{\partial q_j}.
 \eeq
 We remind that we are still identify ${\mathcal A}\simeq \widetilde{\mathcal T}\times  \mathcal{M}$, so that we consider $p_j$ and $q_j$ 
 as the functions on $\widetilde{\mathcal T}\times  \mathcal{M}$,
 $$
 q_j \equiv q_j(\vec{t}, M), \quad p_j \equiv p_j(\vec{t}, M), \quad H_k \equiv H_k\Bigl( q_j(\vec{t}, M), p_j(\vec{t}, M), \vec{t}\Bigl).
 $$
 The compatibility of the system (\ref{action1}) means (see, e.g., \cite{B}) that  all
 $$
 c_{kl} := \{H_k, H_l\} +\frac{\partial H_k}{\partial t_l} - \frac{\partial H_l}{\partial t_k}
 $$
 are the Casimir functions{\footnote{ Warning: here,
 $$
 \frac{\partial H_k}{\partial t_l} =  \frac{\partial }{\partial t_l}\Bigl(H_k(\vec{p}, \vec{q}, \vec{t})\bigl|_{\vec{p}, \vec{q} \equiv \mathrm{const}}\Bigr).
 $$}}
  (maybe depending on the times $t_k$). We shall  assume that
 \beq\label{Casimir}
 c_{kl} = 0 \quad \forall \,\, k,l.
 \eeq
 This assumption works  for all example of the isomonodromic deformations equations that we know.
 The {\it classical action differential} can be defined as the differential form on $\widetilde{\mathcal T}\times  \mathcal{M}$,
 \beq\label{action2}\nonumber
 \omega_{\mbox{cla}}= \sum p_jdq_j - \sum H_kdt_k \equiv 
 \sum_k\left(\sum_j p_j\frac{\partial q_j}{\partial t_k} -  H_k\right)dt_k +  \sum_k\left(\sum_j p_j\frac{\partial q_j}{\partial m_k} \right)dm_k
 \eeq
and, using assumption (\ref{Casimir}), it is easy to check that  it is closed on the trajectories of the dynamical system (\ref{action1}), i.e.,

$$
d_{\mathcal T}\left(\omega_{\mbox{cla}}\bigl|_{M\equiv \mathrm{const}}\right) =0.
$$

 Note that in  those cases when the logarithm of the tau function is the generating function for the
 Hamiltonians $H_k$,  the Jimbo-Miwa-Ueno differential form is
\beq\label{JMU3}\nonumber
 \omega_{\mathrm{JMU}}= \sum H_kdt_k,
 \eeq
so that  the integral (\ref{tau}) is the {\it truncated} action integral,
\beq\label{action3}\nonumber
\ln \tau =  \intop_{\vec{t_1}}^{\vec{t_2}} \sum_k H_kdt_k.
\eeq
Suppose that instead of this integral we need to study the complete action, i.e. the integral,
\beq\label{action4}\nonumber
S\equiv S(\vec{t_1}, \vec{t_2}, M) =  \intop_{\vec{t_1}}^{\vec{t_2}} \omega_{\mbox{cla}}(M) \equiv 
\intop_{\vec{t_1}}^{\vec{t_2}}\sum_k\left(\sum_j p_j\frac{\partial q_j}{\partial t_k} -  H_k\right)dt_k. 
\eeq
Then, the usual variational calculus arguments show that, similar to (\ref{action0}),  in any $m_j$-derivative of $S$ 
the integral terms would disappear. In fact, we have,
$$
\frac{\partial S}{\partial m_{j_0}}= 
\intop_{\vec{t_1}}^{\vec{t_2}}\sum_k\left(\sum_j \frac{\partial p_j}{\partial m_{j_0}}\frac{\partial q_j}{\partial t_k} +
p_j\frac{\partial^2 q_j}{\partial t_k\partial m_{j_0}} -  \frac{\partial H_k}{\partial p_j}
\frac{\partial p_j}{\partial m_{j_0}}-  \frac{\partial H_k}{\partial q_j}
\frac{\partial q_j}{\partial m_{j_0}}\right)dt_k
$$
$$
= \left.\sum_{j,k}p_j\frac{\partial q_j}{\partial m_{j_0}}\right|_{\vec{t_1}}^{\vec{t_2}} +
\intop_{\vec{t_1}}^{\vec{t_2}}\sum_k\left(\sum_j \frac{\partial p_j}{\partial m_{j_0}}\frac{\partial q_j}{\partial t_k} 
- \frac{\partial q_j}{\partial m_{j_0}}\frac{\partial p_j}{\partial t_k}-  \frac{\partial H_k}{\partial p_j}
\frac{\partial p_j}{\partial m_{j_0}}-  \frac{\partial H_k}{\partial q_j}
\frac{\partial q_j}{\partial m_{j_0}}\right)dt_k
$$
\beq\label{action5}
= \left.\sum_{j,k}p_j\frac{\partial q_j}{\partial m_{j_0}}\right|_{\vec{t_1}}^{\vec{t_2}}
\eeq
and the integral term vanishes because of the equations of motion (\ref{action1}). 
Comparison (\ref{action0}) and (\ref{action5}) makes one to suspect some deep connection between  the tau function and the classical action. 
And, indeed, taken the full exterior derivation of $ \omega_{\mbox{cla}}\equiv \omega_{\mbox{cla}}(\vec{t}, M) $, one obtains,
\beq\label{Omegadef}\nonumber
d \omega_{\mathrm{cla}} = \sum_{j}d_{ \mathcal{M}}p_j\wedge d_{ \mathcal{M}}q_j \equiv \Omega.
\eeq
The form $\Omega$ is the symplectic form associated with the dynamical system (\ref{action1}). Note, that both, the form $\Omega$ and the form 
$\Omega_0$ from (\ref{Omega0def}) are the closed 2- form on  $\mathcal{M}$ and they do not depend on the times  $\mathcal{T}$. This observation together with the  similarities  of the variational identities (\ref{action0}) and (\ref{action5}) allow us to formulate the following conjectures.
\vskip .1in
\noindent
{\bf Conjecture 1}. Suppose that the parameter space $\mathcal{A}$ is equipped with the symplectic structure.
Let $\Omega$ be a corresponding symplectic form and $\Omega_0$ be the two-form defined in (\ref{Omega0def}).
Then, there exists a numerical constant $\gamma$ such that,
$$\Omega_0 = \gamma\Omega.$$
\vskip .1in
\noindent
If this conjecture is true then the two 1-forms, $\omega$ and $\omega_{\mbox{cla}}$, coincide up to the total differential. Hence our next conjecture
\vskip .1in
\noindent
{\bf Conjecture 2}. {\it There exists a rational  function $G(\vec{p}, \vec{q}, \vec{t})$ of $\vec{p}, \vec{q}, \vec{t}$ such that,}
\beq\label{conj2}
\omega = \gamma \omega_{\mbox{cla}} + dG(\vec{p}, \vec{q}, \vec{t}).
\eeq
{\it Moreover, the function $G(\vec{p}, \vec{q}, \vec{t})$ is explicitly computable.}
\vskip .1in
As it has already been indicated in the Introduction, the  statement of Conjecture 2 has been proven to be true in the case of the system associated to Schlesinger equations \cite{Malg2}, in the case  of the sine-gordon reduction 
of Painlev\'e III equation \cite{IP}, in the 
case of the (homogenous)  Painlev\'e II equation \cite{ILP}, and in the case of the Painlev\'e I equation \cite{LR}.  In the next section of this paper  we demonstrate  the validity of the both conjectures for the rest of the  Painlev\'e equations, and we also  present, for completeness, the result of \cite{Malg2}.
\begin{rmk}
Restricting (\ref{conj2}) to the  isomonodromic family $ M \equiv$ const, 
one arrives to the identity
$$
\sum_k\frac{\partial \ln \tau}{ \partial t_k}dt_k=  \sum_k\left(\sum_j p_j(\vec{t}, M)\frac{\partial q_j(\vec{t}, M)}{\partial t_k} -  
H_k\Bigl(\vec{p}(\vec{t}, M), \vec{q}(\vec{t}, M), \vec{t}\Bigr)\right)dt_k 
$$
\beq\label{action10}
+  \sum_k\frac{\partial}{\partial t_k }G\Bigl(\vec{p}(\vec{t}, M), \vec{q}(\vec{t}, M), \vec{t}\Bigr)dt_k.
\eeq
and hence,
\beq\label{action11}
\ln \tau (\vec{t_1}, \vec{t_2}, M) = S(\vec{t_1}, \vec{t_2}, M) + \left.G\Bigl(\vec{p}(\vec{t}, M), \vec{q}(\vec{t}, M), \vec{t}\Bigr)\right|_{\vec{t_1}}^{\vec{t_2}}.
\eeq
This, in turn, would produce, taking into account (\ref{action5}),  the following, alternative to (\ref{action0}), formula for the $m_j$ - derivative of
$\ln \tau$,
\beq\label{action12}
\frac{\partial \ln \tau}{\partial m_{j_0}} = \left.\sum_{j,k}p_j\frac{\partial q_j}{\partial m_{j_0}}\right|_{\vec{t_1}}^{\vec{t_2}} + 
\left.\frac{\partial G}{\partial m_{j_0}}\right|_{\vec{t_1}}^{\vec{t_2}}.
\eeq
This version of the variational logarithmic derivatives of the tau functions turns out even more efficient then (\ref{action0}) in the concrete examples related to the ``constant problem''. Indeed, the particular cases of  (\ref{action12}) have been used in \cite{BIP} in evaluation of the constant terms in the
 asymptotics of the several basic  distribution functions of random matrix theory expressible in terms of the Painlev\'e transcendents.
\end{rmk}
\begin{rmk}
In the pioneering papers \cite{ILST} and \cite{ILT13},  the evaluation of the constant pre-factors  has been partially  based on the conjectural  interpretations 
of these constant factors  as the generating
functions of the canonical transformations between the  Darboux asymptotic coordinates associated with the different critical points.  
This generating function interpretation of the constant pre-factors, which in the case of Painlev\'e II, III and I has been  proven in \cite{IP,ILP} and \cite{LR},
can be considered as a direct corollary of (\ref{action11}).  
\end{rmk}

\section{Painlev\'e equations}\label{3}
In this section we present the exact realization of  the relations (\ref{conj2}) -- (\ref{action10})  between the tau-function and the classical action
for all six Painlev\'e equations and also for the case of the system associated to Schlesinger equation. We shall start with the second Painlev\'e  equation which we
take as a case study and illustrate in detail the general constructions of Section \ref{2}. In particular, we will describe exactly the both spaces,
$\mathcal A$ and $\mathcal M$ corresponding to the Painlev\'e II equation.
The other Painlev\'e equations will be treated with less details. We won't describe  the space $\mathcal M$
for other Painlev\'e equations explicitly, and instead we will refer the reader either to \cite{JM} or to Chapter 5 of \cite{FIKN}.  

\subsection{Painlev\'{e} II }
According to \cite{JM}, the second Painlev\'e equation describes the isomonodromic deformations of the
$2\times2$ linear system having only one irregular singular point at $z =\infty$ of the Poincar\'e rank 3,
\beq\label{p2A1}
\dfrac{d\Phi}{dz}=A\left(z\right) \Phi,\qquad 
A\left(z\right)
={A_{2}}{z^2}+ {A_{1}}{z}+
{A_{0}}.
\eeq
Following again \cite{JM}, we normalize the system by the conditions,
$$
\mbox{Tr}\,A(z) \equiv 0, \quad A_2 = \left(\begin{array}{cc}
1&0\\
0&-1
\end{array}\right), \quad A_{1,11} = A_{1,22} =0,
$$
so that the matrix coefficients $A_1$ and $A_0$ can be written in the form,
\beq\label{p2A2}\nonumber
A_1=\left(\begin{array}{cc}
0&k\\
-\frac{2p}{k}&0
\end{array}\right),\quad  A_0=\left(\begin{array}{cc}
p+\frac{t}{2}&-kq\\
-\frac{2}{k}(\theta+pq)&-p-\frac{t}{2}
\end{array}\right),
\eeq
and the complex parameters $p$, $q$, $k$, $\theta$, and $t$ can be taken as the original coordinates on the
corresponding space $\mathcal{A}$,
\beq\label{p2A3}
\mathcal{A} = \left\{ (p, q, k,  \theta, t)\right\}.
\eeq

The formal solution of system (\ref{p2A1}) at its only (irregular) singular point, $z = \infty$, has the structure (cf. (\ref{formalsol})),
$$
\Phi_{\mathrm{form}}(z) \equiv G(z)e^{\Theta(z)}
$$
\beq\label{p2formal}
=\left(I+\dfrac{g_1}{z}+\dfrac{g_2}{z^2}+\dfrac{g_3}{z^3}+\dots\right)e^{\Theta(z)},\quad \Theta(z)=\sigma_3\left(\dfrac{z^3}{3}+\dfrac{t z}{2}-{\theta} \ln z \right),
\eeq
with the first three matrix coefficients $g_k$, $k = 1,2,3$ given as functions on the space $\mathcal{A}$ (\ref{p2A3}) by the explicit
formulae,
\beq\label{p2g1}
g_1=\left(\begin{array}{cc}
-H&-\frac{k}{2}\\
-\frac{p}{k}&H
\end{array}\right),
\eeq
\beq\label{p2g2}
g_2=\left(\begin{array}{cc}
	\frac{H^2}{2}+\frac{p}{4}-\frac{t\theta}{4}&-\frac{kH}{2}+\frac{kq}{2}\\
	\frac{pH}{k}-\frac{pq}{k}-\frac{\theta}{k}&	\frac{H^2}{2}+\frac{p}{4}
	+\frac{t\theta}{4}
	\end{array}\right),
	\eeq
\beq\label{p2g3}
g_3=\left(\begin{array}{cc}
-\frac{H^3}{6}-\frac{Hp}{4}+\frac{Ht}{6}+\frac{Ht\theta}{4}+\frac{pq}{6}+\frac{\theta^2}{6}+\frac{\theta}{3}&-\frac{kH^2}{4}+\frac{kqH}{2}+\frac{kp}{8}+\frac{kt}{4}-\frac{kt\theta}{8}\\
-\frac{pH^2}{2k}+\frac{Hpq}{k}+\frac{H\theta}{k}+\frac{p^2}{4k}+\frac{pt\theta}{4k}+\frac{pt}{2k}&\frac{H^3}{6}+\frac{Hp}{4}-\frac{Ht}{6}+\frac{Ht\theta}{4}-\frac{pq}{6}-\frac{\theta^2}{6}+\frac{\theta}{6}
\end{array}\right),
\eeq
where
\begin{equation}\label{p2H}
{{H}}= \frac{p^2}{2} + pq^2+\dfrac{pt}{2}+ q\theta.
\end{equation}

The system (\ref{p2A1}) has seven canonical solutions, characterized  by the asymptotic condition,
$$
\Phi_{j}(z) \simeq \Phi_{\mathrm{form}}(z), \quad z \rightarrow \infty, \quad \dfrac{(2j-5)\pi}{6}< \arg z<\dfrac{(2j-1)\pi}{6} ,\quad \Phi_7 (z) = \Phi_1(z)e^{-2\pi i\theta\sigma_3}.
$$
and hence it has six Stokes matrices, $S_j = \Phi^{-1}_j(z)\Phi_{j+1}(z)$, which have the following triangular structure (
for more detail see \cite{JM} or Chapter 5 of \cite{FIKN}),
$$
S_{2k+1} = \left(\begin{array}{cc}
1&0\\
s_{2k+1}&1
\end{array}\right),\quad S_{2k} = \left(\begin{array}{cc}
1&s_{2k}\\
0&1
\end{array}\right),
$$
and satisfy one cyclic relation,
$$
S_1S_2...S_6 = e^{-2\pi i\theta\sigma_3}.
$$
Also, as it follows from (\ref{p2formal}),  the parameter $\theta$ determines  the formal monodromy exponent, $\Theta_0 =\theta\sigma_3$. This means, that the space $\mathcal{M}$ in the case of system ({\ref{p2A1})  can be identified with the algebraic variety of dimension 4,
$$
\mathcal M = \left\{ \vec{m} = (s_1, s_2, ..., s_6, \theta): 1+s_1s_2=(1+s_4s_5)e^{2\pi i \theta}, 1+s_2s_3=(1+s_5s_6)e^{-2\pi i\theta}, s_1+s_3+s_1s_2s_3=-s_5 e^{2\pi i\theta}\right\}.
$$
Equation (\ref{p2formal}), also tells us that the parameter $t$ is the only isomonodromic time, so that, in the case of system (\ref{p2A1}) we have,  
$$
\mathcal{T} = \left\{t\right\}.
$$

The isomonodromic deformations of system (\ref{p2A1}), i.e., the conditions,
$$
p = p (t, M),\,\,\,q = q (t, M),\,\,\,k = k (t, M),\quad M \equiv \mbox{const},
$$
yield  the second linear matrix differential equation, this time with respect to $t$, for the function $\Phi(z) \equiv \Phi(z; t, M)$, 
\beq\label{p2B1}
\dfrac{d\Phi}{dt}=B\left(z\right) \Phi,\qquad 
B\left(z\right)
= {B_{1}}{z}+ {B_{0}},
\eeq
where
$$\arraycolsep=1.4pt\def\arraystretch{1.4}
B_1=\dfrac{1}{2}\left(\begin{array}{cc}
1&0\\
0&-1
\end{array}\right),\quad  B_0=\dfrac{1}{2}\left(\begin{array}{cc}
0&k\\
-\frac{2p}{k}&0
\end{array}\right).
$$
Equations (\ref{p2A1}) and (\ref{p2B1}) form a {\it Lax pair} (cf. (\ref{isosys})),
 \beq\nonumber\label{Laxp2}\arraycolsep=4pt\def\arraystretch{2}
\left\{ \begin{array} {l}
 \dfrac{d\Phi}{dz} =  A(z)\Phi,\\
 \dfrac{d\Phi}{dt}=B(z)\Phi,
 \end{array}\right.
 \eeq
whose compatibility condition (\ref{isomeg0}), in terms of the functional parameters $p, q, k$ and $\theta$ reads
\begin{equation}\label{p2}\arraycolsep=4pt\def\arraystretch{2}
\begin{array}{l}
\dfrac{dq}{dt}=p+q^2+\dfrac{t}{2},\\
\dfrac{dp}{dt}=-2pq-\theta,\\
\dfrac{dk}{dt}=-kq,\\
\dfrac{d\theta}{dt} =0.\\
\end{array}
\end{equation}
The last equation of this system is just the statement that $\theta$, as the part of the monodromy data, is constant.
The third  equation gives  $\ln k(t)$ as the antiderivative of $-q(t)$. The first two first order differential
equations are equivalent to one second order differential equation, indeed, the Painlev\'e II equation,
\beq\label{p2equ}
q_{tt}=tq+2q^3+\alpha,\quad \alpha=\dfrac{1}{2}-\theta.
\eeq

Assuming that
$$
\theta \equiv \mbox{const}
$$
one can easily see that the function (\ref{p2H}) is the Hamiltonian of the second Painlev\'e equation (\ref{p2equ}) with respect to the symplectic form,
$$
\Omega = dp\wedge dq,
$$
Indeed, the first and the second  equations in (\ref{p2}) are just 
$$
\frac{dq}{dt} = \frac{\partial H}{\partial p}, \quad\mbox{and}\quad \frac{dp}{dt} = -\frac{\partial H}{\partial q},
$$
respectively.

 Let us now discuss the forms $\omega_{\mathrm{JMU}}$ and  $\omega$ corresponding to (\ref{p2A1} ).
The linear system  (\ref{p2A1}) has only $\infty$ as its singular point. Therefore, the  general definition (\ref{jmu}) of the
form $\omega_{\mathrm{JMU}}$ transforms to the equation,
\begin{equation}\label{jmup2}\nonumber
\omega_{\mathrm{JMU}} = -\operatorname{res}_{z=\infty}\operatorname{Tr}\left(G^{-1}(z)\frac{dG(z)}{dz}\frac{d\Theta(z)}{dt}\right) dt.
\end{equation}
Plugging (\ref{p2formal})  at the right hand side we arrive at the formulae,
$$
\omega_{\mathrm{JMU}} = -\mathrm{Tr}\left(\frac{1}{2}g_1\sigma_3\right),
$$
or, taking into account (\ref{p2g1}) (cf. (\ref{ham0}),
\beq\label{taup21}
\omega_{\mathrm{JMU}} \equiv \frac {d\ln \tau}{dt} dt = Hdt,
\eeq
Similarly, the general definition (\ref{mb})  of the form $\omega$ transforms
to the equation,
\begin{equation}\label{omegap2}
\omega = \operatorname{res}_{z=\infty}\operatorname{Tr}\left(
 A\lb z\rb dG\lb z\rb {G\lb z\rb}^{-1}\right).
\end{equation}
Plugging (\ref{p2formal}) into (\ref{omegap2}) we arrive at the formula,
\beq\label{omegap22}\nonumber
\omega=\mathrm{Tr}\Bigl(A_2dg_3 -A_2dg_2 g_1-A_2dg_1 g_2 +A_2 dg_1 g_1^2+A_1dg_2-A_1dg_1 g_1+A_0d g_1\Bigr).
\eeq
Now, it is more involved to plug (\ref{p2g1}) - (\ref{p2g3}) into the right hand side of the last equation. However, after performing
some algebra, the final expression comes out rather simple, 
\ben
\omega =-\dfrac{1}{3}qdp+\dfrac{2}{3}pdq-\theta \dfrac{dk}{k}+\dfrac{2}{3}tdH-\dfrac{1}{3}Hdt-\dfrac{2\theta-1}{3}d\theta,
\ebn 
and can be in turn easily transformed to the equation, 
\beq\label{omegaclap2}
\omega=pdq-Hdt+d\left(\frac{2}{3}Ht-\frac{1}{3}qp-\theta \ln k -\frac{\theta^2}{3}+\frac{\theta}{3}\right)+\ln k\, d\theta.
\eeq
If we again assume that  
$$
\theta \equiv \mbox{const},
$$
relation (\ref{omegaclap2}) reduces to
\beq\label{omegaclap22}
\omega=pdq-Hdt+d\left(\frac{2}{3}Ht-\frac{1}{3}qp-\theta  \ln k\right).
\eeq

Equation (\ref{omegaclap22}) proves Conjecture 2 in the case of the $2\times2$ system (\ref{p2A1}) with the additional constraint,
$\theta \equiv \mbox{constant}$. Indeed, this  is exactly the formula  (\ref{conj2}) with the specification{\footnote{From the point of view of the asymptotic analysis of the tau functions outlined in the Introduction, the appearance of the non-local term $\ln k=\intop qdt$ is not an obstacle since the functional parameter $k$ equals $-2g_{1,12}$ (see \eqref{p2g1}) where $g_1$ is the first coefficient of the series \eqref{p2formal}. This means that, similar to the functions $p$ and $q$, it can be recovered from the underlining Riemann-Hilbert problem, see also \cite{BBDI}.}}
\beq\label{Gp2}\nonumber
G(p,q, t) = \frac{2}{3}Ht-\frac{1}{3}qp-\theta \ln k.
\eeq
The corresponding equation (\ref{action10}) is
$$
\frac{d\ln \tau}{dt} = p\frac{dq}{dt} - H + \frac{d}{dt}\left(\frac{2}{3}Ht-\frac{1}{3}qp-\theta  \ln k\right) 
$$
\beq\label{action10p2}
=  p\frac{dq}{dt} - H + \frac{d}{dt}\left(\frac{2}{3}Ht-\frac{1}{3}qp\right)+\theta q.
\eeq
We also note that Conjecture 1 follows  directly from  (\ref{omegaclap22}).
\begin{rmk} Together with (\ref{taup21}), equation (\ref{action10p2}) is just an identity which can  be proven directly by
substituting (\ref{taup21}) into the left hand side of  (\ref{action10p2}). However, it would be quite difficult to guess, without 
having the concept of the form $\omega$,  the
existence of such connection between the truncated and the full action differentials. We believe that the fact that in the
case of the Painlev\'e dynamical systems the truncated action differs from the full action by a total differential
is a manifestation  of their Lax-pair  integrability. 
\end{rmk}
\begin{rmk} Let us denote $$
p_1=p,\quad q_1=q, \quad p_2=\ln k,\quad q_2=\theta,
$$
then the whole system \eqref{p2} becomes Hamiltonian with the same Hamiltonian \eqref{p2H},
$$
{{H}}= \frac{p_1^2}{2} + p_1q_1^2+\dfrac{p_1t}{2}+ q_1q_2,
$$
and with respect to  the symplectic form,
$$
\Omega = dp_1\wedge dq_1 +  dp_2\wedge dq_2.
$$
Moreover, (\ref{omegaclap22})
can be written as
\beq\label{omegaclap23}\nonumber
\omega=\sum_{k=1}^{2}p_kdq_k-Hdt+d\left(\frac{2}{3}Ht-\frac{1}{3}q_1p_1-q_2 p_2 -\frac{q_2^2}{3}+\frac{q_2}{3}\right).
\eeq
That is, if we associate with the linear system (\ref{p2A1}) not just the second Painlev\'e
equation (\ref{p2equ}) but the full system \eqref{p2} of the  isomonodromic deformation equations of (\ref{p2A1}), then
the G-function in the relation (\ref{conj2}) will be totally local and, in fact, polynomial in the Darboux coordinates,
$$
G(p_1, p_2, q_1, q_2, t) = \frac{2}{3}Ht-\frac{1}{3}q_1p_1-q_2 p_2 -\frac{q_2^2}{3}+\frac{q_2}{3}.
$$
It is also worth noticing that since $H$ does not contain $p_2$, the variable $q_2$  is an action variable, i.e., it is constant, as
it should be.
\end{rmk}
\begin{rmk} The Lax pair (\ref{p2A1}),  (\ref{p2B1}) is not the only Lax pair for the Painlev\'e II equation. If we use  another Lax pair,
for instance the Lax pair of Flaschka and Newell \cite{FN}, we would get the another tau function, another Hamiltonian and another
form $\omega$. However, the Conjectures 1 and 2 would still be true  - see \cite{ILP} and \cite{BIP}. It is an interesting issue 
how much the Hamiltonian aspects we are promoting depend on the concrete Lax pair realization of the Painlev\'e equations.
\end{rmk}

\subsection{Painlev\'{e} I}
The results of this subsections belong to O. Lisovyy and J. Roussillon, and we follow here  their paper \cite{LR}.
The linear system associated with the first Painlev\'e equation is the $2\times2$  matrix ODE with one irregular singular point
of Poincar\'e rank 5 at $z = \infty$ and with one Fuchsian singular point at $z=0$, 

\beq\label{fuchspvip1}
\dfrac{d\Phi}{dz}=A\left(z\right) \Phi,\qquad 
A\left(z\right)
={A_{4}}{z^4}+ {A_{2}}{z^2}+ {A_{1}}{z}+A_0 +\dfrac{A_{-1}}{z}.
\eeq
The matrix coefficients are,
$$\arraycolsep=1.4pt\def\arraystretch{1.2}
A_4=\left(\begin{array}{cc}
4&0\\
0&-4
\end{array}\right),\quad  A_2=\left(\begin{array}{cc}
0&-4q\\
4q&0
\end{array}\right),\quad A_1=\left(\begin{array}{cc}
0&-2p\\
-2p&0
\end{array}\right),\quad  A_0=\left(\begin{array}{cc}
2q^2+t&-2q^2-t\\
2q^2+t&-2q^2-t
\end{array}\right),\quad A_{-1}=-\frac{1}{2}\left(\begin{array}{cc}
0&1\\
1&0
\end{array}\right),
$$
and so that the  space $\mathcal{A}$ is parametrized by $p$, $q$, and $t$,
$$
\mathcal{A} = \{(p,q,t)\}.
$$

The formal solution at $z = \infty$ is given by the series,
\beq\label{P1infty}
\Phi_{\mathrm{form}}(z)=\left(I+\dfrac{g_1}{z}+\dfrac{g_2}{z^2}+\dfrac{g_3}{z^3}+\dfrac{g_4}{z^4}+\dfrac{g_5}{z^5}+O\left(\dfrac{1}{z^6}\right)\right)e^{\Theta(z)},\quad \Theta(z)=\sigma_3\left(\dfrac{4z^5}{5}+{t z}\right)
\eeq
with the explicit formulae for the first five  coefficient matrices $g_k$  given  by the equations,
\beq\label{p1g1}\arraycolsep=2pt\def\arraystretch{1.5}
g_1=\left(\begin{array}{cc}
-H&0\\
0&H
\end{array}\right),\quad
g_2=\left(\begin{array}{cc}
\frac{H^2}{2}&\frac{q}{2}\\
\frac{q}{2}&	\frac{H^2}{2}
\end{array}\right),\quad 
g_3=\left(\begin{array}{cc}
-\frac{H^3}{6}-\frac{2p-t^2}{24}&\frac{qH}{2}+\frac{p}{4}\\
-\frac{qH}{2}-\frac{p}{4}&\frac{H^3}{6}+\frac{2p-t^2}{24}
\end{array}\right),
\eeq
\beq\label{p1g2}\arraycolsep=3pt\def\arraystretch{1.5}
g_4=\left(\begin{array}{cc}
\frac{H^4}{24}+\frac{2p-t^2}{24}H+\frac{q^2}{8}&\frac{qH^2}{4}+\frac{pH}{4}+\frac{2q^2+t}{8}\\
\frac{qH^2}{4}+\frac{pH}{4}+\frac{2q^2+t}{8}&\frac{H^4}{24}+\frac{2p-t^2}{24}H+\frac{q^2}{8}
\end{array}\right),
\eeq
\beq\label{p1g3}\arraycolsep=3pt\def\arraystretch{1.5} g_5=\left(\begin{array}{cc}
-\frac{H^5}{120}-\frac{2p-t^2}{48}H^2-\frac{5q^2-2t}{40}H-\frac{4pq+1}{160}&\frac{qH^3}{12}+\frac{pH^2}{8}+\frac{2q^2+t}{8}H+\frac{2p-t^2}{48}q+\frac{1}{16}\\
-\frac{qH^3}{12}-\frac{pH^2}{8}-\frac{2q^2+t}{8}H-\frac{2p-t^2}{48}q-\frac{1}{16}&\frac{H^5}{120}+\frac{2p-t^2}{48}H^2+\frac{5q^2-2t}{40}H+\frac{4pq+1}{160}
\end{array}\right).
\eeq
where
\beq\label{hamp1}
H=\frac{p^2}{2}-2q^3-tq.
\eeq
The Fuchsian point $z=0$ is a resonant point and hence the generic theory outlined in the Introduction is not applicable.
In fact, the behavior of the solution $\Phi(z)$ at  $z=0$ is given by the formula \cite{LR},
$$
\Phi(z) = \left(\begin{array}{cc}
1&\frac{1}{2}\\
1&-\frac{1}{2}
\end{array}\right)
z^{-\frac{1}{2}\sigma_3}\hat{\Phi}(z),
$$
where $\hat{\Phi}(z)$ is holomorphic and invertible at $z = 0$.

The set of monodromy data $\mathcal{M}$  of system (\ref{fuchspvip1}) consists of ten Stokes matrices associated with
the irregular singularity at $z = \infty$ out of which, due to the symmetry $z \rightarrow -z$ of the system, only two are in fact free,
i.e. $\dim \mathcal{M} = 2$ (see \cite{LR} for details). The  structure of the essential singularity of $\Phi(z)$ at infinity 
described in (\ref{P1infty}) indicates that the parameter $t$ is the only isomonodromic time.  The corresponding Lax pair is formed by
equation (\ref{fuchspvip1}) and the following additional matrix equation,
\beq\label{fuchspvi1}
\dfrac{d\Phi}{dt}=B\left(z\right) \Phi,\qquad 
B\left(z\right)
= {B_{1}}{z}+ \dfrac{{B_{-1}}}{z},
\eeq
$$\arraycolsep=1.4pt\def\arraystretch{1.4}
B_1=\left(\begin{array}{cc}
1&0\\
0&-1
\end{array}\right),\quad  B_{-1}=\left(\begin{array}{cc}
q&-q\\
q&-q
\end{array}\right).
$$
The compatibility condition  of (\ref{fuchspvip1}) and (\ref{fuchspvi1}) yields the system of ODEs on $q\equiv q(t)$ and $p\equiv p(t)$,

\begin{equation}\label{p1}\arraycolsep=4pt\def\arraystretch{2}
\begin{array}{l}
\dfrac{dq}{dt}=p,\\
\dfrac{dp}{dt}=6q^2+t.\\
\end{array}
\end{equation}
which is equivalent to the Painlev\'e I equation
\beq\label{p11}\nonumber
q_{tt}=6q^2+t.
\end{equation}
The system ({\ref{p2}) is a Hamiltonian system with the Hamiltonian (\ref{hamp1}). 

We are passing now to the forms $\omega_{\mathrm{JMU}}$ and  $\omega$ corresponding to the Painlev\'e I system (\ref{p2}).
Because $z=0$ is the resonant Fuchsian point we strictly speaking can not use the definitions (\ref{jmu}) and (\ref{mb}) for the
forms $\omega_{\mathrm{JMU}}$ and $\omega$. However, following \cite{LR}, we  take (\ref{jmu}) and (\ref{mb}), where the  undefined contribution
of the resonant point $z=0$ simply ignored,  as the definitions of these forms in the case of the Painlev\'e I equations, i.e., as in the case of the
second Painlev\'e equation,  we put
\begin{equation}\label{jmup1}
\omega_{\mathrm{JMU}} = -\operatorname{res}_{z=\infty}\operatorname{Tr}\left(G^{-1}(z)\frac{dG(z)}{dz}\frac{d\Theta(z)}{dt}\right) dt,
\end{equation}
and
\begin{equation}\label{omegap1}
\omega = \operatorname{res}_{z=\infty}\operatorname{Tr}\left(
 A\lb z\rb dG\lb z\rb {G\lb z\rb}^{-1}\right),
\end{equation}
where $G(z)$ and $\Theta(z)$ are the series and the exponent from (\ref{P1infty}). As it is shown in \cite{LR}
such  approach preserves the validity of  Lemmas \ref{lemjmu} and \ref{lemma2}. Just as in the case of  Painev\'e II,
we obtain from (\ref{jmup1}) at once that
\beq\label{taup1}\nonumber
\omega_{\mathrm{JMU}}\equiv \frac{d\ln\tau}{dt} dt=2Hdt.
\eeq
The form $\omega$ needs more work.

Introduce the matrix coefficients $h_j$ of the series inverse to the series (\ref{P1infty}), 
\beq\label{hp1}
\left(I+\dfrac{h_1}{z}+\dfrac{h_2}{z^2}+\dfrac{h_3}{z^3}+\dfrac{h_4}{z^4}+O\left(\dfrac{1}{z^5}\right)\right):=\left(I+\dfrac{g_1}{z}+\dfrac{g_2}{z^2}+\dfrac{g_3}{z^3}+\dfrac{g_4}{z^4}+\dfrac{g_5}{z^5}+O\left(\dfrac{1}{z^6}\right)\right)^{-1}.
\eeq
We have for the first four coefficients the relations,

$$
h_1=-g_1,\quad h_2=-g_2+g_1^2,\quad h_3=-g_3+g_2g_1+g_1g_2-g_1^3,\quad
$$

$$
h_4=-g_4+g_3g_1+g_1g_3+g_2^2-g_2g_1^2-g_1g_2g_1-g_1^2g_2+g_1^4.
$$
Plugging (\ref{hp1}) and (\ref{P1infty})  into (\ref{omegap1}) we arrive at the formula,
$$
\omega=\mathrm{Tr}\Bigl(-A_4(h_4 dg_1+h_3 dg_2+h_2 dg_3+h_1 dg_4+dg_5)-A_2(h_2dg_1+h_1dg_2+dg_3)-A_1(h_1dg_1+dg_2)-A_0dg_1\Bigr).
$$
Using \eqref{p1g1}--\eqref{p1g3} we get after (quite a lot of) simplifications 
$$
\omega=\dfrac{6}{5}pdq-\dfrac{4}{5}qdp-\dfrac{2}{5}Hdt+\dfrac{8}{5}tdH,
$$
and properly combining the terms,
\beq\label{conj2p1}
\omega=2\left[pdq-Hdt+d\left(\frac{4Ht}{5}-\frac{2pq}{5}\right)\right].
\eeq

Equation (\ref{conj2p1}) proves Conjectures 1 and 2, with $\gamma =2$, in the case of the $2\times 2$ system (\ref{fuchspvip1})  and gives the 
explicit formula for $G(p,q,t)$,
\beq\label{Gpqtp1}\nonumber
G(p,q,t) = \frac{2}{5} \Bigl(4Ht - 2pq\Bigr).
\eeq
The corresponding equation (\ref{action10}) is
\beq\label{action10p1}\nonumber
\frac{d\ln\tau}{dt} = 2\left(p\frac{dq}{dt} - H\right)  +  \frac{2}{5}\frac{d}{dt} \Bigl(4Ht - 2pq\Bigr),
\eeq
and, of course, can be easily checked directly.

\subsection{Painlev\'{e} III}

The linear system associated with the third Painlev\'e equation we take again from \cite{JM}. This is the system,
\beq\label{fuchspvi00}
\dfrac{d\Phi}{dz}=A\left(z\right) \Phi,\qquad 
A\left(z\right)
= \frac{A_{-2}}{z^2}+ \frac{A_{-1}}{z}+
{A_{0}},
\eeq
with 
$$\arraycolsep=1.4pt\def\arraystretch{1.2}
A_0=\dfrac{1}{2}\left(\begin{array}{cc}
t&0\\
0&-t
\end{array}\right),\quad  A_{-1}=\left(\begin{array}{cc}
-{\theta_\infty}&-qkt\\
\frac{pq(t-p)}{kt}+\frac{\theta_0+\theta_\infty}{k}-\frac{2\theta_\infty p}{kt}&{\theta_\infty}
\end{array}\right),\quad  A_{-2}=\left(\begin{array}{cc}
p-\frac{t}{2}&-kt\\
\frac{p(p-t)}{kt}&-p+\frac{t}{2}
\end{array}\right).
$$
The system has two irregular  singular   points at $z=\infty$ and $z=0$, both of the Poincar\'e rank 1.
The corresponding formal solutions are:
\beq\label{Php3inf}
\Phi^{(\infty)}_{\mathrm{form}}(z)=\left(I+\dfrac{g_{\infty,1}}{z}+O\left(\dfrac{1}{z^2}\right)\right)e^{\Theta_\infty(z)},\quad \Theta_\infty(z)=\sigma_3\left(\frac{tz}{2}-{\theta_\infty} \ln z \right),
\eeq
with
\beq\label{g1inf}\arraycolsep=3pt\def\arraystretch{2}
g_{\infty,1}=\left(\begin{array}{cc}
-\frac{H}{2}-\frac{pq}{2t}+\frac{\theta_\infty^2-\theta_0^2}{2t}+\frac{t}{2}&{kq}\\
\frac{pq(t-p)}{kt^2}+\frac{\theta_0+\theta_\infty}{kt}-\frac{2\theta_\infty p}{kt^2}&\frac{H}{2}+\frac{pq}{2t}-\frac{\theta_\infty^2-\theta_0^2}{2t}-\frac{t}{2}
\end{array}\right),
\eeq
at $z = \infty$, and 
\begin{equation}\label{p3G}
\Phi^{(0)}_{\mathrm{form}}(z)=G_{0}\left(I+{g_{0,1}}{z}+O\left({z^2}\right)\right)e^{\Theta_0(z)}, \quad \Theta_0(z)=\sigma_3\left(\frac{t}{2z}+{\theta_0} \ln z \right)
\end{equation}
with
\beq\label{g1zero}\arraycolsep=3pt\def\arraystretch{2}
g_{0,1}=\left(\begin{array}{cc}
-\frac{H}{2}-\frac{pq}{2t}-\frac{\theta_\infty^2-\theta_0^2}{2t}+\frac{t}{2}&\frac{qa}{t}\left(p-t\right)+\frac{a}{t}(\theta_\infty-\theta_0)\\
-\frac{1}{ta}\left(pq+\theta_0+\theta_\infty\right)&\frac{H}{2}+\frac{pq}{2t}+\frac{\theta_\infty^2-\theta_0^2}{2t}-\frac{t}{2}
\end{array}\right),
\eeq
at $z=0$. In (\ref{g1inf}) and (\ref{g1zero}), 
\begin{equation}\label{p3H}
H=\dfrac{1}{t}\left(2p^2q^2+p(2t-2tq^2+(4\theta_\infty-1)q)-2tq(\theta_0+\theta_\infty)+\theta_\infty^2-\theta_0^2\right),
\end{equation}
and $G_{0}$ diagonalizes matrix $A_{-2}$,
$$G^{-1}_0A_{-2}G_0=-\dfrac{t\sigma_3}{2},$$
and it is chosen in the form
\beq\label{G0p3}
G_0=\dfrac{1}{\sqrt{k}}\left(\begin{array}{cc}
k&-{k}\\
\frac{p}{t}&\frac{t-p}{t}
\end{array}\right)a^{-\frac{\sigma_3}{2}},
\eeq
with $a$ being an extra gauge parameter, so that the full space $\mathcal{A}$ is  seven  dimensional,
\beq\label{Ap3}
\mathcal{A} = \left\{p, q, k, a, t, \theta_0, \theta_{\infty}\right\}.
\eeq

From the series (\ref{Php3inf}) and (\ref{p3G}) it follows that $\theta_{\infty}$ and $\theta_{0}$ are the formal monodromy exponents
and the parameter $t$ is the isomonodromic time. The isomonodromicity with respect to $t$ yields the second differential equation
for $\Phi(z) \equiv \Phi(z,t)$,
\beq\label{fuchspvi100}
\dfrac{d\Phi}{dt}=B\left(z\right) \Phi,\qquad 
B\left(z\right)
= {B_{1}}{z}+ {B_{0}}+ \frac{B_{-1}}{z},
\eeq
%
where,
$$\arraycolsep=1.4pt\def\arraystretch{1.4}
B_1=\dfrac{1}{2}\left(\begin{array}{cc}
1&0\\
0&-1
\end{array}\right),\quad  B_0=\dfrac{1}{t}\left(\begin{array}{cc}
0&-qkt\\
\frac{pq(t-p)}{kt}+\frac{\theta_0+\theta_\infty}{k}-\frac{2\theta_\infty p}{kt}&0
\end{array}\right),\quad  B_{-1}=\left(\begin{array}{cc}
\frac{t-2p}{2t}&{k}\\
\frac{p(t-p)}{kt^2}&\frac{2p-t}{2t}
\end{array}\right).
$$
The compatibility condition of the matrix equations (\ref{fuchspvi00}) and (\ref{fuchspvi100}) implies the following
dynamical system on (\ref{Ap3}),
\begin{equation}\arraycolsep=4pt\def\arraystretch{2.2}
\label{p3}
\begin{array}{l}
\dfrac{dq}{dt}=\dfrac{4pq^2}{t}-{2q^2}+\dfrac{q(4\theta_\infty-1)}{t}+2,\\
\dfrac{dp}{dt}=-\dfrac{4p^2q}{t}+\dfrac{p(4tq-4\theta_\infty+1)}{t}+2\theta_0+2\theta_\infty,\\
\dfrac{dk}{dt}=-\dfrac{4pqk}{t}+2qk-\dfrac{2\theta_\infty k}{t},\qquad \dfrac{da}{dt}=\dfrac{a}{t}\left(2qt+2\theta_0\right),\\
\dfrac{d\theta_{\infty}}{dt} = 0, \quad \dfrac{d\theta_{0}}{dt} = 0.
\end{array}
\end{equation}
It should be also mentioned,  that the fourth equation, i.e. the equation for the function $a(t)$, follows
from plugging \eqref{p3G} into equation (\ref{fuchspvi100}) --  the second equation  of the Lax pair, and equating
the terms of zero order with respect to $z$.

The last two equations of system (\ref{p3}) just state that $\theta_{\infty}$ and $\theta_0$ as the part of the monodromy
data, are constant. The third and the fourth equations give $\ln k(t)$ and $\ln a(t)$ as the antiderivatives of the simple
combinations of $p$ and $q$. The first two equations are equivalent to the third Painlev\'e equation,
\beq\label{p32}
q_{tt}=\dfrac{(q_t)^2}{q}-\dfrac{q_t}{t}+\dfrac{1}{t}(\alpha q^2+\beta)+ \gamma q^3+\dfrac{\delta}{q},
\eeq
where
$$
\alpha=8\theta_0,\quad \beta=4-8\theta_\infty,\quad \gamma=4,\quad \delta=-4.
$$
Assuming that $\theta_{\infty}$ and $\theta_0$ are numerical constants, the function (\ref{p3H}) becomes  the Hamiltonian of (\ref{p32})
with $p$, $q$ being the canonical variables.  Also, if we denote
\beq\label{darbuxp3}
p_1=p,\quad q_1=q, \quad p_2=\ln k,\quad q_2=\theta_\infty,\quad p_3=\ln a,\quad q_3=\theta_0,
\eeq
then the whole system \eqref{p3} becomes Hamiltonian with the same Hamiltonian \eqref{p3H}, i.e.
with
\begin{equation}\label{p3H2}\nonumber
H=\dfrac{1}{t}\left(2p_1^2q_1^2+p_1(2t-2tq_1^2+(4q_2-1)q_1)-2tq_1(q_2+q_3)+q_2^2-q_3^2\right),
\end{equation}
 and with respect to the symplectic form,
 $$
 \Omega = dp_1\wedge dq_1 + dp_2\wedge dq_2 + dp_3\wedge dq_3.
 $$
 
 The general formulae   (\ref{jmu}) and (\ref{mb}) transform, in the case of system (\ref{fuchspvi00}), into the equations, 
\begin{equation}\label{jmup3}
\omega_{\mathrm{JMU}} = -\operatorname{res}_{z=\infty}
\operatorname{Tr}\left(\Bigl(G^{(\infty)}(z)\Bigr)^{-1}\frac{dG^{(\infty)}(z)}{dz}\frac{d\Theta_{\infty}(z)}{dt}\right) dt
-\operatorname{res}_{z=0}
\operatorname{Tr}\left(\Bigl(G^{(0)}(z)\Bigr)^{-1}\frac{dG^{(0)}(z)}{dz}\frac{d\Theta_{0}(z)}{dt}\right) dt
\end{equation}
and
\begin{equation}\label{omegap3}
\omega = \operatorname{res}_{z=\infty}\operatorname{Tr}\left(
 A\lb z\rb dG^{(\infty)}\lb z\rb {G^{(\infty)} \lb z\rb}^{-1}\right) + 
  \operatorname{res}_{z=0}\operatorname{Tr}\left(
 A\lb z\rb dG^{(0)}\lb z\rb {G^{(0)} \lb z\rb}^{-1}\right),
\end{equation}
respectively. Substituting the series $G^{(\infty, 0)}(z)$ and the exponentials $\Theta_{\infty, 0}(z)$ from
(\ref{Php3inf}) and (\ref{p3G}) into (\ref{jmup3}), we  obtain that
$$
\omega_{\mathrm{JMU}} = 
-\frac{1}{2}\mathrm{Tr}\Bigl(g_{\infty,1}\sigma_3\Bigr)dt -\frac{1}{2}\mathrm{Tr}\Bigl(g_{0,1}\sigma_3\Bigr)dt 
$$
and using (\ref{g1inf}) and (\ref{g1zero}) we arrive at the final formula for $\omega_{\mathrm{JMU}}$,
\beq\label{jmup3final}
\omega_{\mathrm{JMU}} \equiv \frac{d\ln\tau}{dt} dt =  Hdt +\frac{pq}{t}dt -tdt.
\eeq
Note the additional to $Hdt$ terms in the right hand side of (\ref{jmup3final}). Similar substitution of 
$G^{(\infty, 0)}(z)$  from
(\ref{Php3inf}) and (\ref{p3G}) into (\ref{omegap3}) leads us to the formula,
$$
\omega=\mathrm{Tr}\Bigl(A_{-1}d G_0 G_0^{-1}+G_0^{-1}A_{-2}G_0dg_{0,1}-A_0dg_{\infty,1}\Bigr),
$$
and using again (\ref{g1inf}), (\ref{g1zero}) and (\ref{G0p3}) we arrive at the equation,
$$
\omega=pdq+tdH-\theta_\infty\dfrac{dk}{k}-\theta_0\dfrac{da}{a}-tdt.
$$
After regrouping the terms we obtain that
\beq\label{conj2p3}\nonumber
\omega=pdq-Hdt+d\left(Ht-\theta_\infty\ln k-\theta_0\ln a-\frac{t^2}{2}\right)+\ln k\, d{\theta_\infty}+\ln a\,\,\, d\theta_0,
\eeq
or, using the definition (\ref{darbuxp3}) of the canonical coordinates,
\beq\label{conj2p33}
\omega = p_1dq_1 + p_2dq_2 +p_3dq_3-Hdt+d\left(Ht-q_2p_2-q_3p_3-\frac{t^2}{2}\right).
\eeq

Equation (\ref{conj2p33}) proves Conjectures 1 and 2, with $\gamma =1$, in the case of the $2\times 2$ system (\ref{fuchspvi00})  and
gives the 
explicit formula for $G(p_j,q_j,t)$,
\beq\label{Gpqtp3}\nonumber
G(p_1, p_2, p_3,q_1, q_2, q_3,t) = Ht-q_2p_2-q_3p_3-\frac{t^2}{2}.
\eeq
The corresponding equation (\ref{action10}) is
\beq\label{action10p3}
\frac{d\ln\tau}{dt} = p\frac{dq}{dt}-H+\frac{d}{dt}\left(Ht-\theta_\infty\ln k-\theta_0\ln a-\frac{t^2}{2}\right).
\eeq
\noindent
{\it Remark.} One can deduce from \eqref{p3} that
$$
\frac{pq}{t} = \frac{1}{4} \frac{d}{dt}\ln\frac{a}{k} - \frac {\theta_0 + \theta_{\infty}}{2t}.
$$
Combining this with \eqref{jmup3final} and \eqref{action10p3}, we arrive at the equation,
$$
Hdt = pdq - Hdt +d\left(Ht + \frac{1 - 4 \theta_0}{4}\ln k - \frac{1 + 4 \theta_0}{4}\ln a
+\frac{\theta_0 + \theta_{\infty}}{2}\ln t\right),
$$
where $df \equiv d_tf = \frac{df}{dt}$. In other words, although the truncated action, $Hdt$, is not in this case exactly the Jimbo-Miwa-Ueno form
$\omega_{\mathrm{JMU}}$, it still coincides with the full classical action, up to a total differential. 
As we will see, this is true in all other examples when accidentally  $Hdt \neq \omega_{\mathrm{JMU}}$ .
%

%
%
%

%

\subsection{Painlev\'{e} IV}

This time (see again \cite{JM}), the linear system is the $2\times 2$ system with one irregular singular point at $z=\infty$  with the
Poincar\'e rank 2 and one Fuchsian point at $z=0$:
\beq\label{PhiAp4}
\dfrac{d\Phi}{dz}=A\left(z\right) \Phi,\qquad 
A\left(z\right)
= \frac{A_{-1}}{z}+ {A_{0}}+
{A_{1}}z,
\eeq
where
$$\arraycolsep=1.4pt\def\arraystretch{1.2}
A_1=\left(\begin{array}{cc}
1&0\\
0&-1
\end{array}\right),\quad  A_0=\left(\begin{array}{cc}
t&k\\
-\frac{q(4p-q-2t)+4\theta_\infty}{2k}&-t
\end{array}\right),\quad  A_{-1}=\dfrac{1}{2}\left(\begin{array}{cc}
\frac{q(4p-q-2t)}{2}&-{kq}\\
\frac{q^2(4p-q-2t)^2-16\theta_0^2}{4kq}&-\frac{q(4p-q-2t)}{2}
\end{array}\right).
$$
The corresponding formal solution at $z = \infty$ is
\beq\label{Pinfp4}
\Phi_{\mathrm{form}}(z)=\left(I+\dfrac{g_1}{z}+\dfrac{g_2}{z^2}+O\left(\dfrac{1}{z^3}\right)\right)e^{\Theta(z)},\quad  \Theta(z)=\sigma_3\left(\frac{z^2}{2}+t z-\theta_\infty \ln z \right)
\eeq
with 
\beq\label{g1infp4}
g_1=\dfrac{1}{2}\left(\begin{array}{cc}
-\frac{2H+q}{2}&-{k}\\
-\frac{q(4p-q-2t)+4\theta_\infty}{2k}&\frac{2H+q}{2}
\end{array}\right),$$ $$ g_2=\dfrac{1}{8}\left(\begin{array}{cc}
\frac{1}{4}\left((2H+q+2t)^2-4t^2-8\theta_0^2+8\theta_\infty^2\right)&-{k}\left(2H-q-4t\right)\\
\frac{1}{2k}\left((2H-q)(q(4p-q-2t)+4\theta_\infty+4)+		8q\right)&\frac{1}{4}\left((2H+q+2t)^2-4t^2+8\theta_0^2-8\theta_\infty^2\right),
\end{array}\right),
\eeq
and
\beq\label{p4H}
H=2p^2q-\frac{1}{8}q^3-\frac{1}{2}tq^2+\frac{1}{2}(2\theta_\infty-1-t^2)q+2\theta_\infty t-\dfrac{2\theta_0^2}{q}.
\end{equation}
The behavior of the solution of (\ref{PhiAp4}) at the (non-resonant, this time) Fuchsian point  $z =0$ is described by the equation,
\beq\label{Pzerop4}
\Phi^{(0)}(z)=G_0\left(I+O\left(z\right)\right)z^{\theta_0\sigma_3},\quad z\to 0,
\eeq
where $G_0$ diagonalizes the matrix $A_{-1}$,
$$G_0^{-1}A_{-1}G_0=\theta_0 \sigma_3$$
and it is chosen in the form,
\beq\label{Gzerop4}
G_0=\dfrac{1}{2\sqrt{kq\theta_0}}\left(\begin{array}{cc}
-{kq}&-{kq}\\
-\frac{q(4p-q-2t)-4\theta_0}{2}&-\frac{q(4p-q-2t)+4\theta_0}{2}
\end{array}\right)a^{-\frac{\sigma_3}{2}}.
\eeq

The full parameter space, 
\beq\nonumber
\mathcal{A} = \left\{p, q, k, a, t, \theta_0, \theta_{\infty},\right\},
\eeq
is again seven dimensional with $t$ being the isomonodromic time and $\theta_{\infty}$ and 
$\theta_0$ serving as the formal monodromy exponents at the respective singular points. 
The isomonodromicity with respect to $t$ yields the second differential equation for $\Phi(z)$, 
\beq\label{PhiBp4}
\dfrac{d\Phi}{dt}=B\left(z\right) \Phi,\qquad 
B\left(z\right)
= {B_{1}}{z}+ {B_{0}},
\eeq
where
$$\arraycolsep=1.4pt\def\arraystretch{1.4}
  B_1=A_1,\quad B_0=\left(\begin{array}{cc}
0&k\\
\frac{q(4p-q-2t)+4\theta_\infty}{2k}&0
\end{array}\right).
$$
and the compatibility of (\ref{PhiBp4}) and (\ref{PhiAp4}) implies,
\begin{equation}\label{p4}\arraycolsep=1.4pt\def\arraystretch{2}
\begin{array}{l}
\dfrac{dq}{dt}=4pq,\\
\dfrac{dp}{dt}=-2p^2+\frac{3}{8}q^2+qt+\frac{1}{2}t^2-\theta_\infty+\frac{1}{2}-\dfrac{2\theta_0^2}{q^2},\\
\dfrac{dk}{dt}=-(q+2t)k,\quad \dfrac{da}{dt}=\dfrac{4\theta_0}{q}a,\\
\dfrac{d\theta_{\infty}}{dt} = 0, \quad \dfrac{d\theta_{0}}{dt} = 0.
\end{array}
\end{equation}
As in the previous section, the fourth equation follows from the substitution of (\ref{Pzerop4}) into (\ref{PhiBp4}).

Similar to the previous cases, the last equations of (\ref{p4}) manifest the time-independence  of the  formal monodromy exponents, the 
third and the fourth equations express $k$ and $a$ in terms of $p$ and $q$, while the first and the second equations are equivalent to a
 Painlev\'e
equation, this time to the fourth Painlev\'e equation,
\beq\label{p44}
q_{tt}=\dfrac{(q_t)^2}{2q}+\dfrac{3}{2}q^3+4tq^2+{2(t^2-\alpha)}q+\dfrac{\beta}{q},
\eeq
where
$$
\alpha=2\theta_\infty-1,\quad \beta=-8\theta_0^2.
$$
Assuming that $\theta_{\infty}$ and $\theta_0$ are numerical constants, the function (\ref{p4H}) becomes  the Hamiltonian of (\ref{p44})
with $p$, $q$ being the canonical variables.  Also, if we again denote
\beq\label{darbuxp4}
p_1=p,\quad q_1=q, \quad p_2=\ln k,\quad q_2=\theta_\infty,\quad p_3=\ln a,\quad q_3=\theta_0,
\eeq
then the whole system \eqref{p4}   becomes Hamiltonian with the same Hamiltonian \eqref{p4H}, i.e.
with
\begin{equation}\label{p4H2}\nonumber
H=2p_1^2q_1-\frac{1}{8}q_1^3-\frac{1}{2}tq_1^2+\frac{1}{2}(2q_2-1-t^2)q_1+2q_2 t-\dfrac{2q_3^2}{q_1},
\end{equation}
 and with respect to the symplectic form,
 $$
 \Omega = dp_1\wedge dq_1 + dp_2\wedge dq_2 + dp_3\wedge dq_3.
 $$
 
 The general formulae   (\ref{jmu}) and (\ref{mb}) transform, in the case of system (\ref{PhiAp4}), into the equations, 
\begin{equation}\label{jmup4}
\omega_{\mathrm{JMU}} = -\operatorname{res}_{z=\infty}
\operatorname{Tr}\left(\Bigl(G^{(\infty)}(z)\Bigr)^{-1}\frac{dG^{(\infty)}(z)}{dz}\frac{d\Theta_{\infty}(z)}{dt}\right) dt
\end{equation}
and
\begin{equation}\label{omegap4}
\omega = \operatorname{res}_{z=\infty}\operatorname{Tr}\left(
 A\lb z\rb dG^{(\infty)}\lb z\rb {G^{(\infty)} \lb z\rb}^{-1}\right) + 
  \operatorname{res}_{z=0}\operatorname{Tr}\left(
 A\lb z\rb dG^{(0)}\lb z\rb {G^{(0)} \lb z\rb}^{-1}\right),
\end{equation}
respectively. Substituting the series $G^{(\infty)}(z)$ and the exponentials $\Theta_{\infty}(z)$ from
(\ref{Pinfp4}) into (\ref{jmup4}), and using  (\ref{g1infp4}) we  obtain that
\beq\label{jmup4final}
\omega_{\mathrm{JMU}} \equiv \frac{d\ln\tau}{dt} dt =  Hdt +\frac{1}{2}qdt.
\eeq
Note again the additional to $Hdt$ term in the right hand side of (\ref{jmup4final}). Similar substitution of 
$G^{(\infty, 0)}(z)$  from
(\ref{Pinfp4}) and (\ref{Pzerop4}) into (\ref{omegap4}) followed by the  use of (\ref{g1infp4}) and (\ref{Gzerop4}) leads us to the formulae,
$$
\omega=\mathrm{Tr}(A_{-1}dG_0 G_0^{-1}-A_1dg_2+A_1dg_1 g_1-A_0dg_1)
$$
$$
=-\dfrac{1}{2}qdp+\dfrac{1}{2}pdq+\dfrac{1}{2}tdH-\dfrac{1}{2}Hdt-\theta_\infty \dfrac{dk}{k}-\theta_0 \dfrac{da}{a}+\theta_0 d \theta_0-\dfrac{2\theta_\infty-1}{2}d\theta_\infty.
$$
Regrouping the last equation, we arrive at  the final answer for the form $\omega$,
\beq\label{conj2p4}\nonumber
\omega =pdq-Hdt+d\left(\dfrac{Ht}{2}-\dfrac{pq}{2}-\theta_\infty \ln k -\theta_0 \ln a+\dfrac{\theta_0^2}{2}+\dfrac{\theta_\infty}{2}-\dfrac{\theta_\infty^2}{2}\right)
+{\ln k}\,\,d\theta_\infty+\ln a\,\,\, d\theta_0
\eeq
or, using the definition (\ref{darbuxp4}) of the canonical coordinates,
\beq\label{conj2p44}
\omega = p_1dq_1 + p_2dq_2 +p_3dq_3-Hdt+d\left(\frac{Ht}{2}-\frac{p_1q_1}{2} -q_2p_2-q_3p_3 +\frac{q_3^2}{2}
+\frac{q_2}{2} -\frac{q_2^2}{2} \right).
\eeq

Equation (\ref{conj2p44}) proves Conjectures 1 and 2, with $\gamma =1$, in the case of the $2\times 2$ system (\ref{PhiAp4}})  and
gives the 
explicit formula for $G(p_j,q_j,t)$,
\beq\label{Gpqtp4}\nonumber
G(p_1, p_2, p_3,q_1, q_2, q_3,t) = \frac{Ht}{2}-\frac{p_1q_1}{2} -q_2p_2-q_3p_3 +\frac{q_3^2}{2}
+\frac{q_2}{2} -\frac{q_2^2}{2}. 
\eeq
The corresponding equation (\ref{action10}) and the formula for the truncated action are
\beq\nonumber
\frac{d\ln\tau}{dt} = p\frac{dq}{dt}-H+\frac{d}{dt}\left(\dfrac{Ht}{2}-\dfrac{pq}{2}-\theta_\infty \ln k -\theta_0 \ln a\right),
\eeq
and
\beq\nonumber
Hdt = pdq-Hdt+d\left(\dfrac{Ht}{2}-\dfrac{pq}{2}+ \frac{1-2\theta_{\infty}}{2}\ln k -\theta_0 \ln a +\frac{t^2}{2}\right),
\quad  d \equiv d_t,
\eeq
respectively.

\subsection{Painlev\'{e} V}

In \cite{JM}, the following linear system is associated with the fifth Painlev\'e equation, 

\beq\label{PhiAp5}
\dfrac{d\Phi}{dz}=A\left(z\right) \Phi,\qquad 
A\left(z\right)
= A_2+\frac{A_{0}}{z}+ \frac{A_{1}}{z-1},
\eeq
where
$$\arraycolsep=1.4pt\def\arraystretch{1.2}
A_2=\dfrac{t}{2}\left(\begin{array}{cc}
1&0\\
0&-1
\end{array}\right),\quad  A_0=\left(\begin{array}{cc}
-pq-{\theta_\infty}-{\theta_1}&{k}(pq+\theta_\infty+\theta_1-\theta_0)\\
-\frac{1}{k}(pq+\theta_\infty+\theta_1+\theta_0)&pq+{\theta_\infty}+{\theta_1}
\end{array}\right),$$

$$  A_1=\left(\begin{array}{cc}
pq+{\theta_1}&-kq(pq+2\theta_1)\\
\frac{p}{k}&-pq-{\theta_1}
\end{array}\right).
$$
This system has one irregular singular point of Poincar\'e rank 1 at $z =\infty$ and two Fuchsian singular points $z =0$ and $z=1$.
The corresponding formal solution at $z = \infty$ is given by the formulae,
\beq\label{Pinfp5}
\Phi_{\mathrm{form}}(z)=\left(I+\dfrac{g_1}{z}+O\left(\dfrac{1}{z^2}\right)\right)e^{\Theta(z)},\quad z\to \infty, \quad  \Theta(z)=\sigma_3\left(\frac{t z}{2}-{\theta_\infty} \ln z \right)
\eeq
with
\beq\label{g1p5}
g_1=\left(\begin{array}{cc}
-H&\dfrac{k(pq^2-pq+2\theta_1 q -\theta_\infty -\theta_1 +\theta_0)}{t}\\
-\dfrac{2pq-2p+2\theta_\infty+2\theta_1+2\theta_0}{tk}&H
\end{array}\right)
\eeq
and
\begin{equation}\label{p5H}\arraycolsep=4pt\def\arraystretch{2.2}
\begin{array}{c}
H=\dfrac{p^2(q-1)^2q}{t}+p\left(\dfrac{q^2}{t}(\theta_0+3\theta_1+\theta_\infty)+\dfrac{q}{t}(t-2\theta_\infty-4\theta_1)+\dfrac{1}{t}(\theta_\infty+\theta_1-\theta_0)\right)+\dfrac{q2\theta_1}{t}(\theta_\infty+\theta_1+\theta_0)\\
+\dfrac{\theta_0^2-\theta_1^2-\theta_\infty^2+\theta_1 t-2\theta_1\theta_\infty}{t}.\\
\end{array}
\end{equation}
The behavior of the solutions of (\ref{PhiAp5}) at the (non-resonant) Fuchsian points $z=0$ and $z=1$ are described by the equations,
\beq\label{Phizerop5}
\Phi^{(0)}(z)=G_0\left(I+O\left(z\right)\right)z^{\theta_0\sigma_3},\quad z\to 0,
\eeq
and
\beq\label{Phionep5}
\Phi^{(1)}(z)=G_1\left(I+O\left(z-1\right)\right)(z-1)^{\theta_1\sigma_3},\quad z\to 1,
\eeq
respectively. The matrices $G_0$ and $G_1$ diagonalize the matrix coefficients $A_0$ and $A_1$,
$$
G_0^{-1}A_0G_0=\theta_0 \sigma_3,\quad G_1^{-1}A_1G_1=\theta_1 \sigma_3,
$$
and are chosen in the form,
\beq\label{G0p5}
G_0=\dfrac{1}{\sqrt{-4k\theta_0}}\left(\begin{array}{cc}
k(2pq+2\theta_\infty+2\theta_1-2\theta_0)&k\\
2pq+2\theta_\infty+2\theta_1+2\theta_0&1
\end{array}\right)a^{-\frac{\sigma_3}{2}},
\eeq
and
\beq\label{G1p5}
G_1={\dfrac{1}{\sqrt{2k\theta_1}}}\left(\begin{array}{cc}
{k(pq+2\theta_1)}&kq\\
p&1
\end{array}\right)b^{-\frac{\sigma_3}{2}}.
\eeq

The full space 
\beq\label{Ap5}\nonumber
\mathcal{A} = \left\{p, q, k, a, b, t, \theta_0, \theta_1, \theta_{\infty},\right\}.
\eeq
is nine dimensional with $t$ being the isomonodromic time and $\theta_{\infty}$, 
$\theta_0$, and $\theta_1$  serving as  the formal monodromy exponents at the respective singular points. 
The isomonodromicity with respect to $t$ yields the second differential equation for $\Phi(z)$, 
\beq\label{PhiBp5}
\dfrac{d\Phi}{dt}=B\left(z\right) \Phi,\qquad 
B\left(z\right)
= {B_{1}}{z}+ {B_{0}},
\eeq
where
$$\arraycolsep=1.4pt\def\arraystretch{1.4}
B_1=\dfrac{A_2}{t},\quad B_0=\left(\begin{array}{cc}
0&\frac{k}{t}(-pq^2+pq-2\theta_1 q+\theta_\infty+\theta_1-\theta_0)\\
-\frac{1}{tk}(pq+\theta_\infty-p+\theta_1+\theta_0)&0
\end{array}\right),
$$
and the compatibility  of (\ref{PhiBp5}) and (\ref{PhiAp5}) implies,
\begin{equation}\label{p5}\arraycolsep=4pt\def\arraystretch{2.2}
\begin{array}{l}
\dfrac{dq}{dt}=\dfrac{2pq(q-1)^2}{t}+\dfrac{q^2}{t}(\theta_0+3\theta_1+\theta_\infty)+\dfrac{q}{t}(t-2\theta_\infty-4\theta_1)+\dfrac{1}{t}(\theta_\infty+\theta_1-\theta_0),\\
\dfrac{dp}{dt}=-\dfrac{p^2}{t}(3q^2-4q+1)-p\left(\dfrac{2q}{t}(\theta_0+3\theta_1+\theta_\infty)+\dfrac{1}{t}(t-2\theta_\infty-4\theta_1))\right)-\dfrac{2\theta_1}{t}(\theta_\infty+\theta_1+\theta_0),\\
\dfrac{dk}{dt}=-\dfrac{k}{t}(pq^2-2pq+p+2\theta_1 q-2\theta_\infty-2\theta_1),\\
\dfrac{da}{dt}=\dfrac{a}{t}(p-pq^2-2\theta_1 q-2\theta_0),\\
 \dfrac{db}{dt}=-\dfrac{b}{t}(3pq^2+p-4pq+2\theta_\infty q+4\theta_1 q+2\theta_0 q-2\theta_\infty-2\theta_1+t),\\
\dfrac{d\theta_{\infty}}{dt} = 0, \quad \dfrac{d\theta_{0}}{dt} = 0, \quad \dfrac{d\theta_{1}}{dt} = 0.
\end{array}
\end{equation}
As before, the equations for $a$ and $b$ follow from the substitution of (\ref{Phizerop5}) into (\ref{PhiBp5}) and
(\ref{Phionep5}) into (\ref{PhiBp5}), respectively, and they simply express the functions $a(t)$ and $b(t)$ in terms of
$p$ and $q$. The third equation in (\ref{p5}) is also trivial -- just an expression of $k$ in terms of $p$ and $q$,  and
the last three equations are the manifestation of the time-independence of the formal monodromy exponents. The nontrivial
first two equations are equivalent to the fifth Painlev\'e equation,
\beq\label{p55}
q_{tt}=\left(\dfrac{1}{2q}+\dfrac{1}{q-1}\right)(q_t)^2-\dfrac{q_t}{t}+\dfrac{(q-1)^2}{t^2}\left(\alpha q+\dfrac{\beta}{q}\right)+\gamma\dfrac{q}{t}+\delta\dfrac{q(q+1)}{(q-1)},
\eeq
where
$$
\alpha=\dfrac{(\theta_0-\theta_1+\theta_\infty)^2}{2},\quad \beta=-\dfrac{(\theta_0-\theta_1-\theta_\infty)^2}{2},\quad \gamma=(1-2\theta_0-2\theta_1),\quad \delta=-\dfrac{1}{2}.
$$
Assuming that $\theta_{\infty}$, $\theta_0$, and $\theta_1$ are numerical constants, the function (\ref{p5H}) becomes  the Hamiltonian of (\ref{p55})
with $p$, $q$ being the canonical variables.  Also, if we denote
\beq\label{darbuxp5}
p_1=p,\quad q_1=q, \quad p_2=\ln k,\quad q_2=\theta_\infty,\quad p_3=\ln a,\quad q_3=\theta_0,\quad p_4 =\ln b,\quad q_4=\theta_1,
\eeq
then the whole system \eqref{p5}   becomes Hamiltonian with the same Hamiltonian \eqref{p5H}, i.e.
with
\begin{equation}\label{p55H}\arraycolsep=4pt\def\arraystretch{2.2}\nonumber
\begin{array}{c}
H=\dfrac{p_1^2(q_1-1)^2q_1}{t}+p_1\left(\dfrac{q_1^2}{t}(q_3+3q_4+q_2)+\dfrac{q_1}{t}(t-2q_2-4q_4)+\dfrac{1}{t}(q_2+q_4-q_3)\right)+\dfrac{2q_1q_4}{t}(q_2+q_4+q_3)\\
+\dfrac{q_3^2-q_4^2-q_2^2+q_4 t-2q_4q_2}{t},\\
\end{array}
\end{equation}
and with respect to the symplectic form,
 $$
 \Omega = dp_1\wedge dq_1 + dp_2\wedge dq_2 + dp_3\wedge dq_3 +dp_4\wedge dq_4.
 $$
 
 The general formulae   (\ref{jmu}) and (\ref{mb}) transform, in the case of system (\ref{PhiAp5}), into the equations, 
\begin{equation}\label{jmup5}
\omega_{\mathrm{JMU}} = -\operatorname{res}_{z=\infty}
\operatorname{Tr}\left(\Bigl(G^{(\infty)}(z)\Bigr)^{-1}\frac{dG^{(\infty)}(z)}{dz}\frac{d\Theta_{\infty}(z)}{dt}\right) dt
\end{equation}
and
\begin{equation}\label{omegap5}
\omega = \operatorname{res}_{z=\infty}\operatorname{Tr}\left(
 A\lb z\rb dG^{(\infty)}\lb z\rb {G^{(\infty)} \lb z\rb}^{-1}\right) + 
  \operatorname{res}_{z=0}\operatorname{Tr}\left(
 A\lb z\rb dG^{(0)}\lb z\rb {G^{(0)} \lb z\rb}^{-1}\right) 
 \end{equation}
 $$
 +  \operatorname{res}_{z=1}\operatorname{Tr}\left(
 A\lb z\rb dG^{(1)}\lb z\rb {G^{(1)} \lb z\rb}^{-1}\right), 
 $$
respectively. Substituting the series $G^{(\infty)}(z)$ and the exponent $\Theta_{\infty}(z)$ from
(\ref{Pinfp5}) into (\ref{jmup5}) and using (\ref{g1p5}), we  obtain that, similar to the Painlev\'e II case,
\beq\label{jmup5final}
\omega_{\mathrm{JMU}} \equiv \frac{d\ln\tau}{dt} dt =  Hdt. 
\eeq
Substituting  $G^{(\infty, 0, 1)}(z)$  from
(\ref{Pinfp5}),  (\ref{Phizerop5}), and (\ref{Phionep5}) into (\ref{omegap5}) and using  after that
(\ref{g1p5}), (\ref{G0p5}), (\ref{G1p5})  leads us to the formulae,
$$
\omega=\mathrm{Tr}(A_0d G_0 G_0^{-1}+A_1d G_1 G_1^{-1}-A_2d g_1)
$$
$$
=pdq+tdH-\theta_\infty\dfrac{dk}{k}-\theta_0 \dfrac{da}{a}-\theta_1 \dfrac{db}{b}+d\theta_0+d\theta_1.
$$
Regrouping the last equation, we arrive at  the final answer for the form $\omega$,
\beq\label{omegap55}\nonumber
\omega= pdq-Hdt+d\Bigl(Ht-{\theta_\infty}\ln k-\theta_0 \ln a-\theta_1 \ln b +\theta_0+\theta_1\Bigr)
+\ln a\,\, d\theta_0+\ln b \,\,d\theta_1+\ln k \,\,d\theta_\infty,
\eeq
or, using the definition (\ref{darbuxp5}) of the canonical coordinates,
\beq\label{conj2p55}
\omega = p_1dq_1 + p_2dq_2 +p_3dq_3 +p_4dq_4-Hdt
+d\Bigl(Ht-q_2p_2-q_3p_3-q_4p_4 +q_3 + q_4\Bigr).
\eeq

Equation (\ref{conj2p55}) proves Conjectures 1 and 2, with $\gamma =1$, in the case of the $2\times 2$ system (\ref{PhiAp5}})  and
gives the 
explicit formula for $G(p_j,q_j,t)$,
\beq\label{Gpqtp5}\nonumber
G(p_1, p_2, p_3, p_4, q_1, q_2, q_3, q_4, t) = Ht-q_2p_2-q_3p_3-q_4p_4 +q_3 + q_4.
\eeq
The corresponding equation (\ref{action10}) is
\beq\label{action10p5}\nonumber
\frac{d\ln\tau}{dt} = p\frac{dq}{dt}-H+\frac{d}{dt}\Bigl(Ht-{\theta_\infty}\ln k-\theta_0 \ln a-\theta_1 \ln b \Bigr).
\eeq
Again, this equation together with (\ref{jmup5final})  make an identity, which, this time, would not be so easy
to check directly.

\subsection{Painlev\'{e} VI}
Consider  the $2\times 2$ Fuchsian system with 4 regular singularities at $0,1,t$ and $\infty$,

\beq\label{PiAp6}
\dfrac{d\Phi}{dz}=A\left(z\right) \Phi,\qquad 
A\left(z\right)
= \frac{A_{0}}{z}+ \frac{A_{t}}{z-t}+
\frac{A_{1}}{z-1},
\eeq
where
$$A_0, A_1, A_t \in \mathfrak{sl}_2\left(\mathbb C\right),\quad A_0+A_1+A_t=-{\theta_\infty}\sigma_3.$$
Following to \cite{JM}, we introduce the parametrization,
$$
A_0=\left(\begin{array}{cc}
x_0+{\theta_0}&-ux_0\\
\frac{x_0+2\theta_0}{u}&-x_0-{\theta_0}
\end{array}\right),\quad  A_1=\left(\begin{array}{cc}
x_1+{\theta_1}&-vx_1\\
\frac{x_1+2\theta_1}{v}&-x_1-{\theta_1}
\end{array}\right),\quad  A_t=\left(\begin{array}{cc}
x_t+{\theta_t}&-wx_t\\
\frac{x_t+2\theta_t}{w}&-x_t-{\theta_t}
\end{array}\right).
$$
Observe that, $\pm{\theta_0},\pm{\theta_1},\pm{\theta_t}$ are the
eigenvalues of $A_0, A_1, A_t$, and that 
the following constraints are satisfied,
\beq\label{e1}
x_0+{\theta_0}+x_1+{\theta_1}+x_t+{\theta_t}=-{\theta_\infty},
\eeq
\beq\label{e2}
ux_0+vx_1+wx_t=0,
\eeq
\beq\label{e3}
\frac{x_0+2\theta_0}{u}+ \frac{x_1+2\theta_1}{v}+ \frac{x_t+2\theta_t}{w}=0.
\eeq
We also introduce the parameters $k$ and $q$ by  writing  the entry $A_{12}(z)$ of the matrix $A(z)$ as,
\beq
A_{1,2}(z)=\dfrac{k(z-q)}{z(z-1)(z-t)}.\nonumber
\eeq
Notice that, 
\beq\label{e4}
ux_0(1+t)+vx_1+wx_t=k,\quad ux_0t=kq.
\eeq
Finally we put
\beq\label{e5}
p=A_{11}(q)=\frac{x_0+{\theta_0}}{q}+ \frac{x_1+{\theta_1}}{q-1}+ \frac{x_t+{\theta_t}}{q-t}.
\eeq
Solving equations \eqref{e2} and \eqref{e4} with respect to $u,v,w$, we get
\beq\label{uvw}
u=\dfrac{kq}{x_0t},\quad v=\dfrac{k(q-1)}{x_1(1-t)},\quad w=\dfrac{k(t-q)}{x_tt(1-t)}.
\eeq
Next, we express $x_1, x_t$ from \eqref{e1}, \eqref{e5}, and then we express $x_0$ from \eqref{e3}. The result is

$$
x_0=\dfrac{p^2q^2(q-1)(q-t)}{t2\theta_\infty}+\dfrac{pq(q-1)(q-t)}{t}+\dfrac{\theta_{\infty}q(q-t-1)}{2t}+\dfrac{\theta_1^2(t-1)}{2t\theta_\infty(q-1)}-\dfrac{\theta_t^2t(t-1)}{2\theta_\infty(q-t)}
$$
$$
-\dfrac{\theta_1^2(1-t)}{2t\theta_\infty}-\dfrac{\theta_t^2t(t-1)}{2t\theta_\infty}-{\theta_0}-\dfrac{\theta_0^2}{2\theta_\infty},
$$
\beq\label{x0x1xt}
x_1=\dfrac{p^2q(q-1)^2(t-q)}{(t-1)2\theta_\infty}+\dfrac{pq(q-1)(t-q)}{t-1}+\dfrac{\theta_{\infty}(q-1)(t-q-1)}{2(t-1)}-\dfrac{\theta_0^2t}{2q\theta_\infty(t-1)}+\dfrac{\theta_t^2t(t-1)}{2\theta_\infty(q-t)}
\eeq
$$
+\dfrac{\theta_0^2t}{2(t-1)\theta_\infty}+\dfrac{\theta_t^2t(t-1)}{2(t-1)\theta_\infty}-{\theta_1}-\dfrac{\theta_1^2}{2\theta_\infty},
$$
$$
x_t=\dfrac{p^2q(q-1)(t-q)^2}{t(t-1)2\theta_\infty}+\dfrac{pq(q-1)(q-t)}{t(t-1)}+\dfrac{\theta_{\infty}(q-t)(q+t-1)}{2t(t-1)}+\dfrac{\theta_0^2t}{2q\theta_\infty(t-1)}-\dfrac{\theta_1^2(t-1)}{2\theta_\infty t(q-1)}
$$
$$
-\dfrac{\theta_0^2}{2(t-1)\theta_\infty}+\dfrac{\theta_1^2}{2t\theta_\infty}-{\theta_t}-\dfrac{\theta_t^2}{2\theta_\infty}.
$$
Equations (\ref{uvw}) - (\ref{x0x1xt}) provide parametrization of the matrices $A_0$, $A_1$, and $A_t$ by the 
parameters $q, p, k,  \theta_0, \theta_1, \theta_t, \theta_{\infty}$, which will prove to be  the Darboux coordinates,
and by the parameter  $t$ which is the isomonodromic time.

Solutions of (\ref{PiAp6}) have the following behavior at $z = 0,1,t$,  and $\infty$,

$$
\Phi^{(\infty)}(z)=\left(I+O\left(z^{-1}\right)\right)z^{-\theta_\infty\sigma_3},\quad z\to \infty.
$$
$$
\Phi^{(0)}(z)=G_0\left(I+O\left(z\right)\right)z^{\theta_0\sigma_3},\quad z\to 0,
$$
$$
\Phi^{(1)}(z)=G_1\left(I+O\left(z-1\right)\right)(z-1)^{\theta_1\sigma_3},\quad z\to 1,
$$
$$
\Phi^{(t)}(z)=G_t\left(I+g_1(z-t)+O\left( (z-t)^2\right)\right)(z-t)^{\theta_t\sigma_3},\quad z\to t,
$$
where
\beq\label{g1p6}\arraycolsep=1.4pt\def\arraystretch{1.4}
g_1=\left(\begin{array}{cc}
\frac{H}{2\theta_t}-p\frac{q(q-1)}{2\theta_t t(t-1)}-\frac{\theta_{\infty}(q-t)}{2\theta_tt(t-1)}&\frac{Hc}{2\theta_t(1-2\theta_t)}-p\frac{q(q-1)c}{2\theta_t t(t-1)}-\frac{\theta_{\infty}(q-t)c}{2\theta_tt(t-1)}-\frac{\theta_t(2qt-t^2-q)c}{(2\theta_t-1)t(t-1)(q-t)}\\
\frac{H}{2\theta_t(1+2\theta_t)c}+p\frac{q(q-1)}{2\theta_t t(t-1)c}+\frac{\theta_{\infty}(q-t)}{2\theta_tt(t-1)c}-\frac{\theta_t(2qt-t^2-q)}{(2\theta_t+1)t(t-1)(q-t)c}&-\frac{H}{2\theta_t}+p\frac{q(q-1)}{2\theta_t t(t-1)}+\frac{\theta_{\infty}(q-t)}{2\theta_tt(t-1)}
\end{array}\right).
\eeq
and 
\begin{equation}
\label{p6H}
H=p^2\dfrac{q(q-1)(q-t)}{t(t-1)}+p\dfrac{q(q-1)}{t(t-1)}+\dfrac{\theta_\infty(1-\theta_\infty)(q-t)}{t(t-1)}+\dfrac{\theta_0^2(q-t)}{qt(t-1)}-\dfrac{\theta_1^2(q-t)}{(q-1)t(t-1)}+\dfrac{\theta_t^2(t^2-q(2t-1))}{(q-t)t(t-1)}.
\end{equation}
The matrices $G_0$, $G_1$, and $G_t$ diagonalize the matrix residues $A_0$, $A_1$, and $A_t$,
$$
G_0^{-1}A_0G_0=\theta_0\sigma_3,\quad G_1^{-1}A_1G_1=\theta_1\sigma_3,\quad G_t^{-1}A_tG_t=\theta_t\sigma_3.
$$
and they are chosen in the form,
\beq\label{G0p6}
G_0=\sqrt{\dfrac{kq}{2\theta_0t}}\left(\begin{array}{cc}
1&1\\
\frac{1}{u}&\frac{x_0+2\theta_0}{ux_0}
\end{array} \right)a^{-\frac{\sigma_3}{2}},
\eeq
\beq\label{G1p6}
G_1=\sqrt{\dfrac{k(q-1)}{2\theta_1(1-t)}}\left(\begin{array}{cc}
1&1\\
\frac{1}{v}&\frac{x_1+2\theta_1}{vx_1}
\end{array} \right)b^{-\frac{\sigma_3}{2}},
\eeq
\beq\label{Gtp6}
G_t=\sqrt{\dfrac{k(t-q)}{2\theta_tt(1-t)}}\left(\begin{array}{cc}
1&1\\
\frac{1}{w}&\frac{x_t+2\theta_t}{wx_t}
\end{array} \right)c^{-\frac{\sigma_3}{2}}.
\eeq

The whole parameter space $\mathcal{A}$ has dimension 11,
\beq\label{Ap4}
\mathcal{A} = \left\{p, q, k, a, b, c,  t, \theta_0, \theta_1, \theta_t,  \theta_{\infty}\right\}.
\eeq
The isomonodromicity with respect to $t$ yields the second differential equation for $\Phi(z)$,
\beq\label{PhiBp6}
\dfrac{d\Phi}{dt}=-\frac{A_{t}}{z-t} \Phi.
\eeq
and the equations
\beq\label{Gequ}
\dfrac{dG_0}{dt}=\dfrac{A_t}{t}G_0,\quad
\dfrac{dG_1}{dt}=\dfrac{A_t}{t-1}G_1,\quad
\dfrac{dG_t}{dt}=\left(\dfrac{A_0}{t}+\dfrac{A_1}{t-1}\right)G_t,
\eeq
for the gauge matrices $G_0$, $G_1$, $G_t$. The compatibility of (\ref{PhiBp6}) and (\ref{PiAp6})
together with the equations (\ref{Gequ}) imply the following dynamical system on (\ref{Ap4}),
\begin{equation}
\dfrac{dq}{dt}=\dfrac{2pq(q-1)(q-t)}{t(t-1)}+\dfrac{q(q-1)}{t(t-1)},\nonumber 
\end{equation} 
\begin{equation}
\dfrac{dp}{dt}=\dfrac{1}{4t(t-1)}\left( 4p^2(2tq-3q^2-t+2q)+4p(1-2q)+4\theta_\infty(\theta_\infty-1) \right)-\dfrac{\theta_0^2}{q^2(t-1)}+\dfrac{\theta_1^2}{t(q-1)^2}-\dfrac{\theta_t^2}{(q-t)^2},
\nonumber
\end{equation}
\begin{equation}\label{p6}
\dfrac{dk}{dt}=\dfrac{k(2\theta_\infty-1)(q-t)}{t(t-1)},
\end{equation} 
\begin{equation}
\label{p6a}
\dfrac{da}{dt}=-\dfrac{2\theta_0(q-t)a}{qt(t-1)},\quad
\dfrac{db}{dt}=\dfrac{2\theta_1(q-t)b}{t(t-1)(q-1)},\quad
\dfrac{dc}{dt}=\dfrac{2\theta_t(q(2t-1)-t^2)c}{(q-t)t(t-1)},\nonumber
\end{equation}
\begin{equation}\label{pp6}
\frac{d\theta_0}{dt} = \frac{d\theta_1}{dt}=\frac{d\theta_t}{dt}= \frac{d\theta_{\infty}}{dt}=0.
\end{equation}
As before, the only non-trivial equations are the first two, and they are equivalent to the sixth Painlev\'e equation
for the function $q(t)$,
$$
\dfrac{d^2q}{dt^2}=\dfrac{1}{2}\left(\dfrac{1}{q}+\dfrac{1}{q-1}+\dfrac{1}{q-t}\right)\left(\dfrac{dq}{dt}\right)^2-\left(\dfrac{1}{t}+\dfrac{1}{t-1}+\dfrac{1}{q-t}\right)\dfrac{dq}{dt}
$$
\beq\label{p66}
+\dfrac{q(q-1)(q-t)}{t^2(t-1)^2}\left(\alpha+\beta\dfrac{t}{q^2}+\gamma\dfrac{t-1}{(q-1)^2}+\delta\dfrac{t(t-1)}{(q-t)^2}\right),
\eeq
where
$$
\alpha=\dfrac{(2\theta_\infty-1)^2}{2},\quad \beta=-{2\theta_0^2},\quad \gamma={2\theta_1^2},\quad \delta=\dfrac{1-4\theta_t^2}{2}.
$$
Assuming that $\theta_{j}$, $j = 0, 1, t, \infty$  are numerical constants, the function (\ref{p6H}) becomes  the Hamiltonian of (\ref{p66})
with $p$, $q$ being the canonical variables.  Also, if we  denote
$$
p_1=p,\quad q_1=q, \quad p_2={\ln k}{},\quad q_2=\theta_\infty,\quad
 p_3={\ln a}{},\quad q_3=\theta_0,
 $$
\beq\label{darbuxp6}
p_4={\ln b}{},\quad q_4=\theta_1,\quad p_5={\ln c}{},\quad q_5=\theta_t.
\eeq
then the whole system \eqref{p6}--\eqref{pp6}   becomes Hamiltonian with the same Hamiltonian \eqref{p6H}, that is with,
\begin{equation}\nonumber
\label{p6H2}
H=p_1^2\dfrac{q_1(q_1-1)(q_1-t)}{t(t-1)}+p_1\dfrac{q_1(q_1-1)}{t(t-1)}+\dfrac{q_2(1-q_2)(q_1-t)}{t(t-1)}+\dfrac{q_3^2(q_1-t)}{q_1t(t-1)}-\dfrac{q_4^2(q_1-t)}{(q_1-1)t(t-1)}
\end{equation}
$$
+\dfrac{q_5^2(t^2-q_1(2t-1))}{(q_1-t)t(t-1)},
$$
 and with respect to the symplectic form,
 $$
 \Omega = \sum_{j=1}^{5}dp_j\wedge dq_j. 
 $$

The general formulae   (\ref{jmu}) and (\ref{mb}) transform, in the case of system (\ref{PiAp6}), into the equations, 
\begin{equation}\label{jmup6}
\omega_{\mathrm{JMU}} = -\operatorname{res}_{z=t}
\operatorname{Tr}\left(\Bigl(G^{(t)}(z)\Bigr)^{-1}\frac{dG^{(t)}(z)}{dz}\frac{d\Theta_{t}(z)}{dt}\right) dt
\end{equation}
and
\begin{equation}\label{omegap6}
\omega = \operatorname{res}_{z=\infty}\operatorname{Tr}\left(
 A\lb z\rb dG^{(\infty)}\lb z\rb {G^{(\infty)} \lb z\rb}^{-1}\right) + 
  \operatorname{res}_{z=0}\operatorname{Tr}\left(
 A\lb z\rb dG^{(0)}\lb z\rb {G^{(0)} \lb z\rb}^{-1}\right)
\end{equation}
$$
 \operatorname{res}_{z=1}\operatorname{Tr}\left(
 A\lb z\rb dG^{(1)}\lb z\rb {G^{(1)} \lb z\rb}^{-1}\right) + 
  \operatorname{res}_{z=t}\operatorname{Tr}\left(
 A\lb z\rb dG^{(t)}\lb z\rb {G^{(t)} \lb z\rb}^{-1}\right).
 $$
From (\ref{jmup6}) it follows that
$$
\omega_{\mathrm{JMU}} = \theta_t\mbox{Tr}\Bigl(g_1\sigma_3\Bigr),
$$
and taking into account (\ref{g1p6}), we obtain that,
\beq\label{jmup66}\nonumber
\omega_{\mathrm{JMU}}\equiv \frac{d\ln\tau}{dt}dt=Hdt-p\dfrac{q(q-1)}{t(t-1)}dt-\dfrac{\theta_{\infty}(q-t)}{t(t-1)}dt.
\eeq
Similarly, (\ref{omegap6}) reduces to the equation,
$$
\omega=\mathrm{Tr\left( G_0^{-1}A_0dG_0+G_1^{-1}A_1dG_1+G_t^{-1}A_tdG_t-A_tG_tg_1G_t^{-1}dt\right)},
$$
which after  using (\ref{G0p6})-(\ref{Gtp6}) and  simplifying yields  the formula
$$
\omega=pdq-Hdt-\theta_\infty \dfrac{dk}{k}-\theta_0\dfrac{da}{a}-\theta_1 \dfrac{db}{b}-\theta_t \dfrac{dc}{c}+d\theta_\infty.
$$
This, in turn,  can be rewritten as
\beq\label{omegap6final}\nonumber
\omega =pdq-Hdt+d\Bigl({\theta_\infty}{}-{\theta_0}{}\ln a-{\theta_1}{}\ln b-{\theta_t}{}\ln c-{\theta_\infty}{}\ln k\Bigr)
+{\ln k}{}d\theta_\infty+{\ln a}{}d\theta_0
+{\ln b}{}\,d\theta_1+{\ln c}{}\,d\theta_t.
\eeq
or, remembering the definitions  (\ref{darbuxp6}) of the canonical coordinates,
\beq\label{conj2p66}
\omega =\sum_{j=1}^{5}pdq-Hdt+d\Bigl(q_2-q_3p_3-q_4p_4-q_5p_5-q_2p_2\Bigr).
\eeq

Equation (\ref{conj2p66}) proves Conjectures 1 and 2, with $\gamma =1$, in the case of the $2\times 2$ system (\ref{PiAp6})  and
gives the 
explicit formula for $G(p_j,q_j,t)$,
\beq\label{Gpqtp6}\nonumber
G(p_1, p_2, p_3, p_4, p_5, q_1, q_2, q_3, q_4, q_5, t) = q_2-q_3p_3-q_4p_4-q_5p_5-q_2p_2.
\eeq
The corresponding equation (\ref{action10}) and the truncated action are
\beq\nonumber
\frac{d\ln\tau}{dt} = p\frac{dq}{dt}-H-\frac{d}{dt}\Bigl({\theta_0}{}\ln a+{\theta_1}{}\ln b+{\theta_t}{}\ln c+{\theta_\infty}{}\ln k\Bigr),
\eeq
and
\beq\nonumber
Hdt = pdq-Hdt+ d\Bigl(\dfrac{1}{2}\ln\left(\dfrac{k(q-t)}{t(t-1)}\right)-\theta_0 \ln a-\theta_1 \ln b -\theta_t \ln c -\theta_\infty \ln k\Bigr),
\quad  d \equiv d_t,
\eeq
respectively.

\subsection{The system associated to Schlesinger equations}
%
%
This section reproduces the result of \cite{Malg2}  (Subsection  5.6, Remark 5.5).
Once again, we are grateful to M. Mazzocco for informing us about this part of Malgrange's work. 

Consider the Fuchsian system
\beq\label{fuch}
\dfrac{d\Phi}{dz}=A(z)\Phi(z),\quad A(z)=\sum_{\nu=1}^{n}\dfrac{A_\nu}{z-a_\nu}, \quad A_{\nu}\in \mathfrak{sl}_N\left(\mathbb C\right).
\eeq
We assume  that all matrix coefficients $A_{\nu}$ are diagonalizable
\beq\label{A_nu}
A_{\nu}=G_{\nu}\Theta_{\nu}G_{\nu}^{-1};\quad \Theta_{\nu}=\operatorname{diag}\left\{\theta_{\nu,1},\ldots \theta_{\nu,N}\right\},
\eeq
and that their eigenvalues are distinct and non-resonant: 
$$
\theta_{\nu,\alpha}\neq \theta_{\nu,\beta} \mod \mathbb{Z}.
$$
%
%
We also assume that  the residue of $A(z)$ at $z = \infty$ is diagonal, i.e. 
$$
A_{\infty}=-\sum_{\nu=1}^n A_{\nu} = \Theta_{\infty}.
$$
\noindent Solutions of (\ref{fuch}) have the following behavior at the singular points
$$
\Phi^{(\infty)}(z)=(I+O(z^{-1}))z^{-\Theta_\infty},\quad z\to \infty,
$$
$$
\Phi^{(\nu)}(z)=G_\nu(I+g_{\nu,1}(z-a_\nu)+O((z-a_\nu)^2)) (z-a_\nu)^{\Theta_\nu}C_\nu,\quad z\to a_\nu.
$$
The isomonodromic times are now positions of the singular points $a_{\nu}$.  The isomonodromic deformations
with respect to these times yields the equation,
$$
\dfrac{d\Phi}{da_\nu}=B_\nu(z)\Phi(z),\quad B_\nu(z)=-\dfrac{A_\nu}{z-a_\nu}.
$$
The compatibility conditions give the Schlesinger system
\beq\label{shles000}
\dfrac{dA_\mu}{da_\nu}=\dfrac{[A_\mu,A_\nu]}{a_\mu-a_\nu},\quad \mu\neq \nu,
\quad \quad\quad
\dfrac{dA_\nu}{da_\nu}=-\sum_{\mu\neq\nu}\dfrac{[A_\mu,A_\nu]}{a_\mu-a_\nu},
\eeq
and also the equations (cf. equations (\ref{Gequ})),
\beq\label{Geqshles}
\dfrac{dG_\mu}{da_\nu}=\dfrac{A_\nu}{a_\nu-a_\mu}G_\mu,\quad\mu\neq\nu,
\quad \quad\quad\dfrac{dG_\nu}{da_\nu}=-\sum_{\mu\neq\nu}\dfrac{A_\mu}{a_\mu-a_\nu}G_\nu. 
\eeq
Moreover, equations \eqref{Geqshles} imply Schlesinger system \eqref{shles000} for $A_\nu$ given by \eqref{A_nu}.

Following \cite{JMMS} we introduce matrix functions 
\beq\label{darbouxshles}\nonumber
Q_\nu=G_\nu\Theta_\nu,\quad P_\nu=G_\nu^{-1}
\eeq
%
and Hamiltonians
$$
H_\nu=\sum_{\mu\neq\nu}^n\,\dfrac{\operatorname{Tr}(Q_\mu P_\mu Q_\nu P_\nu)}{a_\nu-a_\mu}.
$$
Notice that $A_{\nu} = Q_{\nu}P_{\nu}$. Then, as it is shown in \cite{Malg2}, the isomonodromic equations (\ref{Geqshles}) imply the following Hamiltonian system, which we call the system associated to Schlesinger equations
\beq \label{eq:ass-to-schles}
\dfrac{dP_{\mu,jk}}{da_\nu}=-\dfrac{\partial H_\nu}{\partial Q_{\mu,kj}},\quad \dfrac{dQ_{\mu,jk}}{da_\nu}=\dfrac{\partial H_\nu}{\partial P_{\mu,kj}}.
\eeq
We consider it on the extended space $$\mathcal{A}^{ext}=\{P_{\nu,jk},\,Q_{\nu,jk}:\, \nu=1\ldots n;\, j,k=1\ldots  N\}.$$ It possesses the preserving quantities $ P_{\nu}Q_{\nu}=\Theta_\nu$. Moreover, by a rather  straightforward  calculation, one can show that the general formulae  (\ref{jmu}) and (\ref{mb})
in the case of the Fuchsian system (\ref{fuch}) produce the following expressions of the forms $\omega_{\mathrm{JMU}}$
and $\omega$,
\beq\label{jmushles}\nonumber
\omega_{\mathrm{JMU}}=\sum_{\nu=1}^nH_\nu da_\nu,
\eeq
and 
\beq\label{conj2shles}\nonumber
\omega=\sum_{\nu=1}^n\mathrm{Tr(P_\nu d Q_\nu)}-H_\nu da_\nu.
\eeq
In other words, we have validity of Conjectures 1 and 2 , with $\gamma=1$, in the case of the system (\ref{eq:ass-to-schles}) considered on the space $\mathcal{A}^{ext}$.
Moreover, the form $\omega$ just coincides with $\omega_{\mathrm{cla}}$.

\section{Appendix.}
\subsection{The proof of Lemma \ref{lemjmu}}\label{appendix1}
The proof of Lemma \ref{lemjmu} is rather short. Indeed,  noticing that
	\begin{equation} \label{Gz} 
\left(G^{(\nu)}\right) ^{-1}\dfrac{dG^{(\nu)}}{dz}=\left(G^{(\nu)}\right)^{-1}AG^{(\nu)} -\dfrac{d\Theta_{\nu}}{ dz}
\end{equation}
and plugging this into the right have side of (\ref{jmu}), we have,
 \beq\label{ap1}
\omega_{\mathrm{JMU}}=-\sum_{k=1}^L\sum_{\nu = 1,\ldots, n, \infty} \operatorname{res}_{z=a_\nu} 
\mathrm{Tr}\left(\left( G^{(\nu)}\right) ^{-1}AG^{(\nu)}\, \dfrac{\partial\Theta_{\nu}}{ \partial t_k}\right)dt_k+\sum_{k=1}^L\sum_{\nu = 1,\ldots, n, \infty} \operatorname{res}_{z=a_\nu} \mathrm{Tr}\left(\dfrac{d\Theta_{\nu}}{ dz}\, \dfrac{\partial\Theta_{\nu}}{ \partial t_k}\right)dt_k.
\eeq
The expression $\dfrac{d\Theta_{\nu}}{ dz}\, \dfrac{\partial\Theta_{\nu}}{ \partial t_k}$ has poles of order at least 2, so it does not have residues
and hence the second sum in (\ref{ap1}) vanishes. We also  have,
\begin{equation} \label{Gt}
\dfrac{\partial\Theta_{\nu}}{\partial t_k}=\left(G^{(\nu)}\right)^{-1}B_kG^{(\nu)}-\left(G^{(\nu)}\right)^{-1}\dfrac{\partial G^{(\nu)}}{\partial t_k}.
\end{equation} 
Substituting (\ref{Gt}) into (\ref{ap1}) we transform it to the equation,
$$
 \omega_{\mathrm{JMU}}=-\sum_{k=1}^L\sum_{\nu = 1,\ldots, n, \infty} \operatorname{res}_{z=a_\nu} 
\mathrm{Tr}\Bigl(AB_k\Bigr)dt_k+
\sum_{k=1}^L\sum_{\nu = 1,\ldots, n, \infty} \operatorname{res}_{z=a_\nu} 
\mathrm{Tr}\left(A \dfrac{\partial G^{(\nu)}}{\partial t_k}\Bigl(G^{(\nu)}\Bigr) ^{-1}\right)dt_k.
$$
The function $AB_k$ is rational, therefore sum of its residues is zero. So we get \eqref{ojmu2}.
\subsection{The proof of Lemma \ref{lemma2}}\label{appendix2}

	Denote
	$$
	I=	\sum_{k=1}^L\sum_{j=1}^d\sum_{\nu=1,\ldots,n,\infty}\operatorname{res}_{z=a_{\nu}}\mathrm{Tr}\left(	
	\dfrac{\partial}{\partial m_j}\left(A\dfrac{\partial G^{(\nu)}}{\partial t_k} \left(G^{(\nu)}\right)^{-1}\right)-	
	\dfrac{\partial}{\partial t_k}\left(A\dfrac{\partial G^{(\nu)}}{\partial m_j} \left(G^{(\nu)}\right)^{-1}\right)\right).
	$$
	We have
	$$
I=\sum_{k=1}^L\sum_{j=1}^d\sum_{\nu=1,\ldots,n,\infty}\operatorname{res}_{z=a_{\nu}}\mathrm{Tr}\left(	
\dfrac{\partial A}{\partial m_j}\dfrac{\partial G^{(\nu)}}{\partial t_k} \left(G^{(\nu)}\right)^{-1}-A\dfrac{\partial G^{(\nu)}}{\partial t_k}\left(G^{(\nu)}\right)^{-1}\dfrac{\partial G^{(\nu)}}{\partial m_j}\left(G^{(\nu)}\right)^{-1}	-\dfrac{\partial A}{\partial t_k}\dfrac{\partial G^{(\nu)}}{\partial m_j} \left(G^{(\nu)}\right)^{-1}\right)
	$$
	$$
	+\sum_{k=1}^L\sum_{j=1}^d\sum_{\nu=1,\ldots,n,\infty}\operatorname{res}_{z=a_{\nu}}\mathrm{Tr}\left(	
	A\dfrac{\partial G^{(\nu)}}{\partial m_j}\left(G^{(\nu)}\right)^{-1}\dfrac{\partial G^{(\nu)}}{\partial t_k}\left(G^{(\nu)}\right)^{-1}\right).
	$$
	We use the formula \eqref{Gt} to get rid of $\dfrac{\partial G^{(\nu)}}{\partial t_k}$ and equation \eqref{isomeg0} 
	to replace $\dfrac{\partial A}{\partial t_k}$. Omitting terms with zero residue and after some cancellations we have
	$$
	I=\sum_{k=1}^L\sum_{j=1}^d\sum_{\nu=1,\ldots,n,\infty}\operatorname{res}_{z=a_{\nu}}\mathrm{Tr}
	\left(-\dfrac{\partial A}{\partial m_j}G^{(\nu)}\dfrac{\partial \Theta_\nu}{\partial t_k}\left(G^{(\nu)}\right)^{-1}
	+AG^{(\nu)}\dfrac{d\partial \Theta_\nu}{\partial t_k}\left(G^{(\nu)}\right)^{-1}\dfrac{\partial G^{(\nu)}}{\partial m_j}\left(G^{(\nu)}\right)^{-1}-\dfrac{dB_k}{dz}\dfrac{\partial G^{(\nu)}}{\partial m_j}\left(G^{(\nu)}\right)^{-1} \right)
	$$
$$
+\sum_{k=1}^L\sum_{j=1}^d\sum_{\nu=1,\ldots,n,\infty}\operatorname{res}_{z=a_{\nu}}\mathrm{Tr}\left(-A\dfrac{d\partial G^{(\nu)}}{\partial m_j}
\dfrac{\partial \Theta_\nu}{\partial t_k}\left(G^{(\nu)}\right)^{-1} \right).
$$
We replace $B_k$ using again formula \eqref{Gt}. After that, we notice that the residue of the derivative with respect to z of formal series is zero. Therefore we can "integrate by parts", moving the derivative from one term to another. We do that with the term, where we replaced $B_k$. Using \eqref{Gz}, we have

$$
	I=\sum_{k=1}^L\sum_{j=1}^d\sum_{\nu=1,\ldots,n,\infty}\operatorname{res}_{z=a_{\nu}}\mathrm{Tr}
	\left(-\dfrac{\partial A}{\partial m_j}G^{(\nu)}\dfrac{\partial \Theta_\nu}{\partial t_k}\left(G^{(\nu)}\right)^{-1}+AG^{(\nu)}
	\dfrac{\partial \Theta_\nu}{\partial t_k}\left(G^{(\nu)}\right)^{-1}\dfrac{d G^{(\nu)}}{\partial m_j}\left(G^{(\nu)}\right)^{-1}-A\dfrac{\partial G^{(\nu)}}{\partial m_j}
	\dfrac{\partial \Theta_\nu}{\partial t_k}\left(G^{(\nu)}\right)^{-1} \right)
$$
  $$
  +\sum_{k=1}^L\sum_{j=1}^d\sum_{\nu=1,\ldots,n,\infty}\operatorname{res}_{z=a_{\nu}}\mathrm{Tr}
  \left(-G^{(\nu)}\dfrac{\partial \Theta_\nu}{\partial t_k}\left(G^{(\nu)}\right)^{-1}\dfrac{\partial G^{(\nu)}}{\partial m_j}
  \left(G^{(\nu)}\right)^{-1}A+\dfrac{\partial \Theta_\nu}{\partial t_k}\left(G^{(\nu)}\right)^{-1}\dfrac{\partial G^{(\nu)}}{\partial m_j}
  \dfrac{d \Theta_\nu}{dz} \right)
  $$
  $$
   +\sum_{k=1}^L\sum_{j=1}^d\sum_{\nu=1,\ldots,n,\infty}\operatorname{res}_{z=a_{\nu}}\mathrm{Tr}
   \left(G^{(\nu)}\dfrac{\partial \Theta_\nu}{\partial t_k}\left(G^{(\nu)}\right)^{-1}\dfrac{\partial^2 G^{(\nu)}}{\partial z\partial m_j}\left(G^{(\nu)}\right)^{-1}\right).
  $$
  Finally using \eqref{Gz} one more time we get $I=0$.
\vspace{0.5cm}

\noindent
{\small {\bf Acknowledgements}. 
	In addition to M. Mazzocco, we would also like to thank P. Boalch and O. Lisovyy for many very useful discussions and comments.
The present work  was supported by the  NSF Grant DMS-1700261 and Russian Science Foundation grant No.17-11-01126.}

\end{document}